\documentclass[
reprint,
superscriptaddress,
nofootinbib,
aps, pra
]{revtex4-1}
\usepackage{subcaption}
\usepackage{float}
\usepackage{amsmath, amsfonts}
\usepackage{soul}
\usepackage{bm}	
\usepackage{bbm}				 % Blackboard type letters
\usepackage{braket}				 % Allows Dirac notation
\usepackage{mathtools}			 % Math capabilities 
\usepackage{graphicx} 			 % Allows for insertion of figures
\usepackage{pst-plot}			 % Allows for post script pictures
\usepackage[hidelinks]{hyperref} % Hyper reference
\usepackage{color}
\usepackage{CJKutf8}
\newcommand{\nn}{\nonumber}

\begin{document}

%========================================
\title{Zero-dimensional models for gravitational and scalar QED decoherence}

\author{Qidong Xu}
\email[]{qidong.xu.gr@dartmouth.edu}
\affiliation{Department of Physics and Astronomy, Dartmouth College, Hanover, New Hampshire 03755, USA}

\author{M. P. Blencowe}
\email[]{miles.p.blencowe@dartmouth.edu}
\affiliation{Department of Physics and Astronomy, Dartmouth College, Hanover, New Hampshire 03755, USA}

\date{\today}

\begin{abstract}
We investigate the dynamics of two quantum mechanical oscillator system-bath toy models obtained by truncating  to zero spatial dimensions linearized gravity coupled to a massive scalar field and scalar QED. The scalar-gravity toy model maps onto the phase damped oscillator, while the scalar QED toy model approximately maps onto an oscillator system subject to two-photon damping. The toy models provide potentially useful insights into solving for open system quantum dynamics relevant to the full scalar QED and weak gravitational field systems, in particular operational probes of the decoherence for initial scalar field system superposition states.          
\end{abstract}

\maketitle
%========================================

%========================================
%========================================
\section{Introduction}
\label{Introduction}
The non-existence of macroscopic mass system quantum superposition states under everyday conditions is commonly understood to be due to interactions with the system environment; air molecules, photons, and internal system defects cause the rapid decoherence of position and energy superposition states into apparent mixtures of either/or alternatives that are indistinguishable from a classical statistical distribution \cite{caldeira1983,joos1985,zurek2003}. By placing the system in ultrahigh vacuum, shielding it from external electromagnetic radiation, and cooling the system down to its ground state, quantum mechanics would in principle allow for macroscopic system superposition states to be prepared and measured. However, there is one environment that cannot be screened out--gravity, as expressed dynamically at the classical level through Einstein's field equations \cite{blencowe2013,anastopoulos2013,oniga2016,bassi2017,delisle2019,asprea2021,anastopoulos2021}. 

From a fundamental perspective, it is interesting to try to quantify the effect of the gravitational environment on macroscopic mass/energy superposition states; even if the predicted gravitationally induced decoherence times are much longer than for everyday, electromagnetic environments, having a good quantitative understanding of the former would allow us to place in principle, unavoidable bounds on the coherence times of macroscopic superposition states, and furthermore help point the way towards possible future experiments to probe the role of gravity in enforcing macroscopic classicality.

Under terrestrial or space-based laboratory conditions corresponding to weak spacetime curvature \cite{kaltenbaek2016}, it should be sufficient to work with linearized gravity \cite{donoghue1994}, where the matter system-gravitational environment action is quadratic in metric deviations $h_{\mu\nu}$ from Minkowski spacetime $\eta_{\mu\nu}[={\mathrm{diag}}(-+++)]$: $g_{\mu\nu}=\eta_{\mu\nu}+\kappa h_{\mu\nu}$, where $\kappa=\sqrt{32\pi G}$ (with natural units $\hbar=c=1$). Furthermore, modeling the matter system through quantum excitations of a massive scalar field $\phi$, a ``first-principles" starting point for investigating gravitational decoherence is the following action: 
\begin{equation}
S[\phi,h_{\mu\nu}]=S_M[\phi]+S_E[h_{\mu\nu}]+S_I[\phi,h_{\mu\nu}],
\label{sysenveq}
\end{equation} 
where the system, environment, and interaction actions are respectively:
\begin{equation}
S_M[\phi]=-\frac{1}{2}\int d^4 x \left(\eta^{\mu\nu} \partial_{\mu}\phi\partial_{\nu}\phi+m^2\phi^2\right),
\label{scalarSeq}
\end{equation}
\begin{align}
S_E[h_{\mu\nu}]=&\int d^4 x\left(-\frac{1}{2} \partial^{\rho} h^{\mu\nu} \partial_{\rho} h_{\mu\nu}+\partial_{\nu} h^{\mu\nu}\partial^{\rho}h_{\mu\rho}\right.\cr &\left.-\partial_{\mu}h \partial_{\nu}h^{\mu\nu}+\frac{1}{2}\partial^{\mu}h\partial_{\mu}h\right),
\label{gravenveq}
\end{align}
and
\begin{equation}
S_{{I}}=\frac{\kappa}{2}\int d^4 x T^{\mu\nu}(\phi)h_{\mu\nu},
\label{inteq}
\end{equation}
with $T^{\mu\nu}(\phi)$ the scalar field energy-momentum tensor, $U_{\mu\nu\rho\sigma}(\phi)$ a quadratic in $\phi$ tensor \cite{arteaga2004}, and $h=h^{\mu}_{\mu}$. 

Quantization might then proceed  through the derivation of a master equation for the density matrix of the scalar matter system, with the (assumed for simplicity) thermal gravitational environmental degrees of freedom traced out \cite{blencowe2013,anastopoulos2013,oniga2016}. Alternatively, a quantum Langevin equation might be derived for the scalar matter field operator, again with the gravitational environment integrated out. One route to obtaining such effective equations is the closed time path integral approach, which is particularly suited to field systems \cite{calzetta2008}. 

However, as a coupled system-environment field theory with a non-quadratic interaction and a gauge symmetry (i.e., general coordinate invariance), the derivation of the quantum gravitational decoherence dynamics presents additional challenges beyond the usual system-environment models considered in non-relativistic quantum mechanics \cite{carmichael1999,petruccione2002,gardiner2004}. One  challenge involves the necessity for making various approximations in order to solve for the reduced system dynamics. For example, in the usual open quantum systems analyses, it is assumed that the system$+$environment is initially in a product state, e.g., the system is in a superposition of two distinct wavepacket or energy states and the environment is in a thermal state. Such a product state can result in an initial ``burst" of decoherence that depends on the upper cut-off physics of the environment, which in the case of gravity is unknown. Furthermore, a Born and possibly Markovian approximation is made \cite{carmichael1999,petruccione2002,gardiner2004}, where the influence of the environment on the system is treated perturbatively to lowest non-trivial order, while the environment is assumed to respond rapidly relative to the system dynamics timescale. 

Another challenge concerns requiring gauge invariance of the calculated decoherence rates for them to be meaningful, in particular when assuming a finite temperature environment that comprises gauge degrees of freedom (e.g., photons or gravitons) \cite{petruccione2002}.  A common, direct approach \cite{carmichael1999,petruccione2002,gardiner2004} to obtaining decoherence rates for open quantum systems, either with or without gauge degrees of freedom, is to examine the time evolution of the off-diagonal matrix elements of the system density operator in the state basis of interest (e.g., energy eigenstates, position eigenstates etc.). However, the density operator is not a gauge invariant quantity. 

A more consistent approach is to extract the decoherence rates through an operational procedure, i.e., involving an in principle measurement that can be ascribed to a particular expectation value of an observable. One such example is the particle detection number density in an atom interference set-up. A minimal way to get scalar matter field quanta initially in superposition states to interfere is by spatially trapping the field quanta in a three-dimensional harmonic confining potential \cite{oniga2017}; the system action (\ref{scalarSeq}) is then supplemented by the term
\begin{equation}
    -\frac{1}{2}\int d^4x m^2\Omega^2 r^2 \phi^2,
    \label{trapeq}
\end{equation}
where $\Omega$ is the characteristic oscillation frequency and the potential center coincides with the spatial origin ${\mathbf{r}}=0$ in the rest frame of the confining potential.
\begin{figure}[t]
\begin{center}
\includegraphics[width=3in]{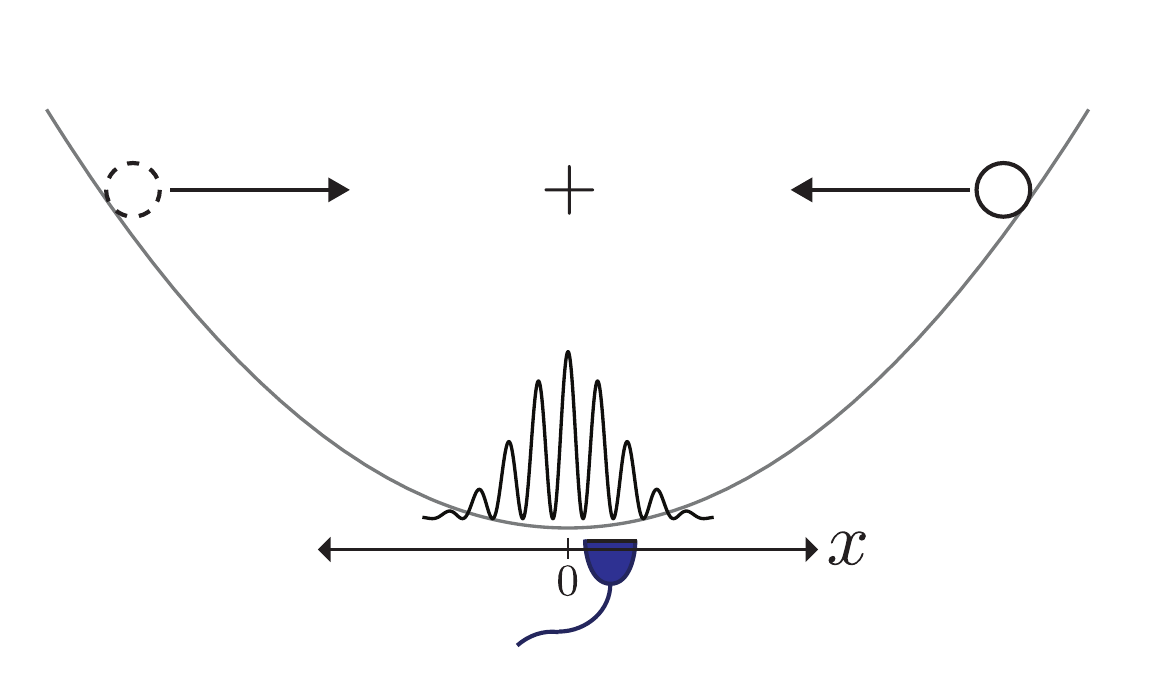} 
\caption{\label{decohereschemefig}Scheme for operationally defining gravitational decoherence. An initial spatial superposition of $N$ nucleon-oscillator coherent states gives rises to a spatial interference pattern in the particle detection probability whenever the particle wavefunctions pass through each other at $x=0$. The particle detector is indicated centered at some given $x$-location. A suppression of the $x$-dependent interference pattern in the particle detection probability is interpreted as gravitational decoherence.}
\end{center}
\end{figure} 
Referring to Fig.~\ref{decohereschemefig}, we might then consider a thought experiment where an initial ($t=0$) $N$-nucleon state  corresponding to being in a superposition of two collective coherent states with coordinate parameters ${\mathbf{r}}=(x,0,0)$,  $x=x_{01}>0,\, x_{0 2}<0$ (and hence superposition separation $x_{0 1}-x_{0 2}$) undergoes gravitational decoherence. Once during every oscillation period, the coherent states in the superposition pass through each other in the region centered at $x=0$, resulting in an interference pattern for the $x$-dependence of the local particle detection probability as indicated in the figure. A measure of the degree of coherence is the so-called ``interferometric visibility", defined below in Eq. (\ref{visibilityeq}); a reduction in visibility over time is interpreted as a signature of gravitationally induced decoherence or dephasing.

In particular,  suppose that we have a particle detector with center of mass worldline $(t,{\bf{r}})$ in the vicinity of ${\mathbf{r}}=0$ and described by the local field observable  $\left(V^{-1}\int_V d{\bf{r}}\phi({\bf{r}},0)\right)^2$, where $\phi({\bf{r}},0)$ is the scalar field operator  (in the Schr\"{o}dinger picture), and $V$ is a coordinate averaging volume (assumed very small) that reflects the fact that a real detector is not pointlike, but rather occupies some nonzero volume in space. The visibility can then be obtained in terms of the following expectation value:
\begin{eqnarray}
&&{\mathrm{Tr}}\left[\rho(t) \left(V^{-1}\int_V d{\bf{r}}\phi({\bf{r}},0)\right)^2  \right]\cr
&&=V^{-2}\int_V d{\bf{r}}d{}{\bf r}'{\mathrm{Tr}}\left[\rho(t)\phi({\bf{r}},0)\phi({\bf{r}}',0)\right],
\label{detectoraveq}
\end{eqnarray}
where the density matrix $\rho(t)$ describes the $N$ nucleons initially in a coherent superposition state and interacting with a thermal graviton bath environment. The expectation value (\ref{detectoraveq}) gives a measure of the spatial particle number density (smeared over the small volume $V$) and is to be viewed as the field-theoretic counterpart to the configuration space probability  density $V^{-1} \int_V d{\bf{r}}\langle {\mathbf{r}}|\rho_{\mathrm{ho}}(t)|{\mathbf{r}}\rangle$ for a single, non-relativistic quantum harmonic oscillator described by the evolving density matrix $\rho_{\mathrm{ho}}(t)$.

The just described set-up shares features of atom and molecular wave interferometry experiments \cite{kovachy2015,xu2019,fein2019}, but utilizing optical traps \cite{li2011,romero2011,bose2017,delic2020}. The latter enables the two wavefunction components making up the superposition to interfere multiple times as they oscillate through each other, rather than just once as in most matter wave interferometry set-ups. Furthermore, no  additional coupled degrees of freedom  such as spins manipulated by external magnetic fields in a Stern-Gerlach-type apparatus \cite{bose2017} are required in  the system-environment action order to implement the interferometer; a full,  relativistic quantum field theoretic analysis can be applied with just the addition of a harmonic confining potential for the scalar field. We must emphasize however that our set-up should not be viewed necessarily as a possible way to feasibly measure gravitational decoherence, but rather as an in-principle operational procedure to quantify the decoherence via the above-defined visibility measure.

With the above-describe challenges in mind, in the present paper we shall consider as a first step, two toy system-environment models that are in turn closely related through dimensional reduction to the above scalar field-gravity system and to scalar field quantum electrodynamics (QED) \cite{boyanovsky1998}. The Lagrangian for scalar QED is
\begin{equation}
L = - (D_{\mu} \phi)^*(D^{\mu} \phi) - m^2 \left(1+\Omega^2 r^2\right) \phi^* \phi - \frac{1}{4} F_{\mu \nu} F^{\mu \nu},
\label{qedlageq}
\end{equation}
where $\phi$ is a complex-valued scalar field, $D_{\mu} = \partial_{\mu} - ie A_{\mu}$ is the covariant derivative, and $F_{\mu\nu}=\partial_{\mu}A_{\nu}-\partial_{\nu}A_{\mu}$ is the electromagnetic field strength tensor. We have also included a three-dimensional harmonic confining potential [c.f. Eq. (\ref{trapeq})] in order to facilitate operational probes of (de)coherence for initial scalar field spatial quantum superposition states as discussed above. 

The models presented in Sec. \ref{0dmodelssec} below are ``toys" in the sense that there is no spatial coordinate--just a time coordinate--and hence are formally zero-dimensional (0d) field models. Our motivation here is to utilize the toy models in order to validate the above-described operational interferometric approach to decoherence as well as certain standard approximation methods, thus giving confidence in eventually applying a similar approach to quantifying actual gravitationally induced decoherence;
the 0d model was in fact utilized in Ref. \onlinecite{blencowe2013} to lend support for an initial gravitational decoherence derivation. 

As zero-dimensional field systems, the toy models lack any gauge symmetry, however. It is for this reason that full scalar field QED is also useful for investigating decoherence and verifying that the considered decoherence measures are gauge invariant. In particular, what constitutes a gauge invariant observable is conceptually clearer in scalar QED than in gravity and thus the former also serves as a useful pedagogical stepping stone towards addressing gravitational decoherence. 

In Sec. \ref{0dmodelssec}, we introduce the 0d toy model Lagrangians. Section \ref{gravsec} analyzes the quantum dynamics of the scalar-weak gravity toy model, by utilizing an exact solution to the full system-environment Schr\"{o}dinger equation assuming an initial system-environment product state, with the oscillator system state expressed in a number state basis and environment in a thermal state. These solutions are then utilized to determine the decoherence dynamics of initial superpositions of system oscillator coherent states through an operational interference fringe visibility analysis that is the single particle counterpart to that described above.  Section \ref{qedsec} analyzes both the classical and quantum dynamics of the scalar QED toy model. In particular, both classical and quantum Langevin equations as well as a quantum master equation are derived for the system oscillator interacting with its oscillator bath. By making various approximations, the 0d model is shown to map onto that of a simpler oscillator system with `two-photon' damping. The master equation is numerically solved to determine the decoherence dynamics of initial superpositions of system oscillator coherent states, again utilizing the operational interference fringe visibility approach. Section \ref{conclusionsec} gives some concluding remarks.     
%========================================
%========================================
\section{0d toy models}
\label{0dmodelssec}
We consider in turn two distinct oscillator system-environment models described by the following Lagrangians:
\begin{align}
L_{\mathrm{grav}}=&\frac{1}{2}M\dot{x}^2 -\frac{1}{2}M\Omega^2 x^2+\sum_i \left(\frac{1}{2}m \dot{q}_i^2-\frac{1}{2} m \omega_i^2 q_i^2\right)\cr
&- \left(\frac{1}{2}M\dot{x}^2 +\frac{1}{2}M\Omega^2 x^2\right)\sum_i \lambda_iq_i
\label{0dgraveq}
\end{align}
and
\begin{align}
L_{\mathrm{qed}} =& \frac{1}{2}M\left(\frac{d}{dt}+ \sum_i \lambda_i q_i\right) x \left(\frac{d}{dt} + \sum_i \lambda_i q_i\right)x  \nn \\
&- \frac{1}{2}M \Omega^2 x^2+ \sum_i\left(\frac{1}{2}m \dot q_i^2 - \frac{1}{2} m \omega_i^2 q_i^2\right).
\label{0dqedeq}
\end{align}
Both model Lagrangians describe an oscillator system with mass $M$ and bare frequency $\Omega$ that is coupled to a bath of independent oscillators with assumed identical masses $m$ and frequencies $\omega_i$. The two models differ in their system-bath couplings; in particular, the system oscillator couples via its energy  to the bath oscillator coordinates in Lagrangian $L_{\mathrm{grav}}$, a 0d analogue of the $T^{\mu\nu}h_{\mu\nu}$ coupling term in Eq. (\ref{inteq}).  On the other hand, the interaction term in Lagrangian $L_{\mathrm{qed}}$ is obtained via a 0d analogue of the gauge principle of minimal coupling: $\partial_{\mu}\rightarrow \partial_{\mu}-i e A_{\mu}$. Expanding out the kinetic energy part of Lagrangian (\ref{0dqedeq}) gives both cubic and quartic interaction terms, which are respectively linear and quadratic in the bath coordinates [c.f. Eq. (\ref{scalar QED L})]; the full, scalar QED Lagrangian (\ref{qedlageq}) possesses analogous nonlinear terms. Note that the coupling strength parameters $\lambda_i$ in Eqs. (\ref{0dgraveq}) and (\ref{0dqedeq}) have different dimensions.

While Lagrangian (\ref{0dgraveq}) is obtained by truncating the spatial dimension from three down to zero for the quantum field theory description of the graviton-matter interaction, and we hence adopt Lagrangian (\ref{0dgraveq})  as our 0d gravitional decoherence model in this work, we note in passing that the Lagrangian also yields the standard Hamiltonian of an optomechanical system under the conditions of weak system-bath coupling, where a single optical mode furnishes the system oscillator degree of freedom, while the bath comprises a very large number of mechanical degrees of freedom. This is in contrast to usually-considered optomechanical systems 
\cite{aspelmeyer2014}, where only one or a few mechanical degrees of freedom are considered. %In the present case, Lagrangian (\ref{0dgraveq}) might describe the dynamics of a light mode of a cavity embedded within a large elastic crystal, or alternatively a light mode trapped between oppositely facing cavity mirrors and coupled via light pressure to a thin, elastic dielectric membrane with large transverse extent \cite{thompson2008}.
With our focus being on the in-principle operational approach to gravitational and scalar QED decoherence in this work, we shall neglect other possible decoherence mechanisms for both 0D models. The optomechanical interpretation and experimental realizations of Eq.~(\ref{0dgraveq}) are explored in a separate companion work \cite{xu2021}. As we shall see later in Sec. \ref{gravsec}, when placed in an initial superposition of coherent states, such a system mode undergoes dephasing without damping--the latter behavior a consequence of the fact that the interaction Hamiltonian commutes with the system Hamiltonian. The resulting, effective system dynamics coincides with that of the so-called `phase damped' oscillator \cite{gardiner2004}. 
%\begin{figure}
%\begin{center}
%\includegraphics[width=3in]{optomech.pdf} 
%\caption{\label{OM}Optomechanical scheme where a cavity-trapped light mode (system)  between oppositely facing mirrors interacts via light pressure with a thin, transversely vibrating dielectric membrane with large transverse extent (bath).}
%\end{center}
%\end{figure}

In Sec. \ref{qedsec}, we show that Lagrangian (\ref{0dqedeq}) describes approximately  an oscillator system subject to two-photon damping \cite{gilles1993}. This is in contrast to the usual quantum Brownian oscillator model with single photon damping and results in qualitatively different decoherence dynamics from the latter for initial superpositions of coherent states.

\section{Scalar gravity model}
\label{gravsec}

\subsection{Solving the model}
\label{gravitysec}
Starting with the Lagrangian $L_{\mathrm{grav}}$ (\ref{0dgraveq}), the system and bath momentum coordinates are
\begin{align}
p &= \frac{\partial L }{\partial \dot x} = M \dot x \left(1- \sum_i \lambda_i q_i\right), \\
p_i &= \frac{\partial L}{\partial \dot q_i} = m \dot q_i,
\end{align}
where we omit the `grav' subscript from now on.
The model Hamiltonian is
\begin{align}
H  =&\frac{p^2}{2M}\left(1- \sum_i \lambda_i q_i\right)^{-1} + \frac{1}{2} M\Omega^2 x^2 \left(1+ \sum_i \lambda_i q_i\right)\nn\\
&+\sum_i\left( \frac{p_i^2}{2 m} 
+ \frac{1}{2} m \omega_i^2 q_i^2\right).
\label{gravhameq}
\end{align}
For weak system-environment (bath) coupling as in the case of weak field gravity, we can Taylor expand the kinetic energy coupled bath term to obtain approximately
\begin{align}
H  =&\left(\frac{p^2}{2M} + \frac{1}{2} M\Omega^2 x^2\right) \left(1+ \sum_i \lambda_i q_i\right)\nn\\
&+\sum_i\left( \frac{p_i^2}{2 m} 
+ \frac{1}{2} m \omega_i^2 q_i^2\right).
\label{gravham2eq}
\end{align}
Quantizing and expressing the Hamiltonian (\ref{gravham2eq}) in terms of the oscillator system and bath  creation and annihilation operators,  which are defined through the respective relations
$x = \sqrt{\frac{\hbar}{2M\Omega}} (a + a^{\dag})$,  $p = i \sqrt{\frac{M\Omega\hbar }{2}}(a^\dag -a)$, $q_i = \sqrt{\frac{\hbar}{2m\omega_i}} (a_i + a_i^{\dag})$, $p_i = i \sqrt{\frac{m\omega_i\hbar }{2}}(a_i^\dag -a_i)$, the scalar gravity model Hamiltonian is
\begin{align}
H  =& \hbar \Omega \left(a^\dag a +\frac{1}{2}\right) \left(1+ \sum_i \lambda_i \left(a_i +a_i^\dag\right)  \right) \nn \\
+& \sum_i \hbar \omega_i \left(a_i^\dag a_i +\frac{1}{2} \right),
\label{classical GR Hamiltonian}
\end{align}
where the system-bath coupling is redefined as $\lambda_i=\sqrt{\frac{\hbar}{2m\omega_i}}\lambda_i$. We recognize in Eq. (\ref{classical GR Hamiltonian}) the familiar form of the standard optomechanical Hamiltonian, but with a bath of mechanical oscillator modes (labelled by index $i$) in contrast to the usually considered situation of just one mechanical mode \cite{aspelmeyer2014}.

Solving for the quantum evolution, we will make the common assumption that the system and bath are in an initial product state $\rho_{s} \otimes \rho_{\mathrm{bath}}$. While the latter assumption facilitates solving for the quantum dynamics, it is not always justified experimentally, since it necessarily requires that the system quantum state can be sufficiently isolated and prepared quickly enough compared to the interaction time scale with the bath degrees of freedom. While such an approximation may be justified for an electromagnetic environment under certain conditions, a mass-energy system can never be isolated from its gravitational environment. Nevertheless, as we shall see, the ability to solve exactly for the scalar gravity model quantum dynamics will give insights into the consequences of assuming a product state. 

It is convenient to work in a basis of system number states and bath coherent states $|n,\{\alpha_i\}\rangle$;  the time evolution for such a state can be written as:
\begin{align}
e^{-\frac{i H t}{\hbar}}|n,\{\alpha_{i}\} \rangle&= \exp\Bigg(-\frac{it}{\hbar} \Big[ \hbar \Omega \left(n+\frac{1}{2}\right)\big(1+ \sum_i \lambda_i (a_{i} \nn \\
+a_{i}^\dag ) \big) +\sum_i \hbar & \omega_i\left(a_{i}^\dag a_{i} +  \frac{1}{2}\right)\Big] \Bigg)\left|n,\{\alpha_{i}\}\right\rangle.
\label{GR evolution} 
\end{align}
Following the analysis of Ref. \cite{bose1997}, 
Eq.~(\ref{GR evolution}) can be evaluated as:
\begin{align}
&e^{-\frac{iH t}{\hbar}}|n,\{\alpha_i\} \rangle = \nn\\
&\exp\Bigg(-it\left[\Omega\left(n+\frac{1}{2}\right) + \sum_i \left(\frac{\omega_i}{2} - \frac{\Omega^2\lambda_i^2 \left(n+\frac{1}{2}\right)^2}{\omega_i} \right) \right]\nn\\
&-\sum_i \frac{i\left(n+\frac{1}{2}\right)^2 \lambda_i^2 \Omega^2}{\omega_i^2} \sin \omega_i t + \frac{1}{2}\sum_i \frac{\lambda_i}{\omega_i}\left(n+\frac{1}{2}\right)\Omega \nn \\
&\times \left[\alpha_i^* \left(1-e^{i\omega_i t}\right) -\alpha_i \left(1-e^{-i \omega_i t}\right) \right]\Bigg) \nn\\
&\times \left| n, \left\{\alpha_i e^{-i\omega_i t} - \frac{\Omega \left(n+\frac{1}{2}\right)}{\omega_i}\lambda_i \left(1-e^{-i\omega_i t}\right) \right\} \right\rangle.
\label{ncoheq}
\end{align}
Supposing the bath to initially be in a thermal state, we can express its initial density matrix in the coherent basis as follows \cite{bose1999scheme}:
\begin{align}
\rho_{\mathrm{bath}} =& \prod_ i \frac{1}{\pi\left(e^{\beta \hbar \omega_i} - 1\right)} \int d \alpha_i^2 \exp\Big(-|\alpha_i|^2 \nn \\
\times& \left(e^{\beta \hbar \omega_i}-1\right) \Big)  |\alpha_i\rangle \langle \alpha_i|,
\label{rhothermeq}
\end{align}
where $\beta^{-1}=k_BT$, with $k_B$ Boltzmann's constant and $T$ the bath temperature. 
Decomposing the initial system-environment state in the number state basis:
\begin{align}
\rho_{\mathrm{initial}} = \sum_{n, n'} C_{n n'}|n\rangle \langle n'| \otimes \rho_{\mathrm{bath}},
\label{initstatedeceq}
\end{align}
we have for the time evolution of the number state outer products after tracing out the bath degrees of freedom:
\begin{align}
&|n(t)\rangle \langle n'(t)| = |n\rangle \langle n'| \exp\Bigg(-it \Big[\Omega (n-n') 
+ \sum_i \frac{\Omega^2 \lambda_i^2}{\omega_i}\nn \\
\times &(n'+n+1)(n'-n) \Big] + i\sum_i \frac{\lambda_i^2 \Omega^2}{\omega_i^2}\sin( \omega_i t)(n'+n+1) \nn \\
\times &(n'-n)-2\sum_i \left(\frac{\Omega \lambda_i(n-n')}{\omega_i} \right)^2 \coth\left( \frac{\beta \hbar \omega_i}{2}\right)\nn \\
\times& \sin^2\left(\frac{\omega_i t}{2} \right)
\Bigg).
\label{outerprodeq}
\end{align}
As we shall show later below in Sec. \ref{gravdecohsec}, the time evolution of an arbitrary initial reduced oscillator system state can be obtained by decomposing in terms of the number state outer product solutions (\ref{outerprodeq}).   
 
In order to carry out the sum over bath degrees of freedom in Eq. (\ref{outerprodeq}), we will assume an ``Ohmic'' bath spectral density with exponential cut-off frequency $\omega_c$:
\begin{align}
\pi \sum_i \lambda_i^2  \delta\left(\omega-\omega_i\right) = C \omega e^{-{\omega}/{\omega_c}}.
\label{bathspeceq}
\end{align}
We note that the bath spectral density is in general determined by the microscopic coupling strength $\lambda_i$ and the spectral structure of the bath modes; here we consider the linear in $\omega$ Ohmic dependence since a thermal graviton bath is Ohmic \cite{blencowe2013}; a cut-off is necessary in order to tame infinities that would otherwise arise for $\omega\rightarrow\infty$ in the intermediate stages of the analysis; the exponential cut-off form is primarily motivated by calculational convenience, enabling exact solutions to be obtained for the oscillator system reduced dynamics in the number state basis. %While the short distance, cut-off physics is in principle known for a concrete material system realisation such as for the vibrating membrane environment \cite{xu2021}, the corresponding short distance, 
In the gravitational decoherence settings, the cut-off energy scale is related to the nature of the `Planckian' physics, which is not known yet. From the perspective of the effective field theory approach to quantum gravity, this unknown high energy physics is assumed to decouple from the low energy dynamics, appearing only in renormalized parameters that are determined phenomenologically \cite{donoghue1994}; as we shall see below, the cut-off dependence can be absorbed in part through a frequency and non-linear Kerr-type self-interaction renormalization. However, the cut-off does affect the initial decoherence dynamics, which is tied to the artificial initial product state assumption.

Using Eq. (\ref{bathspeceq}) to replace the sum in Eq. (\ref{outerprodeq}) with an integral over the continuous variable $\omega$, we obtain
\begin{align}
&|n(t)\rangle \langle n'(t)|= |n\rangle \langle n'| \exp\Bigg( -i\Omega(n-n')t + i\frac{C \Omega ^2}{\pi}\nn \\
& \times(n-n') (n+n'+1)  \left[ \omega_c t - \tan^{-1}(\omega_c t) \right]- \frac{2C\Omega^2}{\pi}
\nn \\
&\times(n-n')^2   \int_{0}^\infty d\omega \omega \coth \left(\frac{\beta \hbar \omega}{2} \right)\frac{\sin^2(\frac{\omega t}{2})}{\omega^2} e^{-{\omega}/{\omega_c}}\Bigg).
\label{outerprod2eq}
\end{align}
Note that for an optomechanical system realization, the environment would have a finite extent resulting in a non-zero, lower frequency cut-off $\omega_1$: $0<\omega_1\ll\omega_c$; here we set the lower frequency cut-off $\omega_1=0$, reflecting the actual gravitational environment with effectively infinite spatial extent. In the case of a non-zero lower frequency cut-off, the system will not fully decohere in the long time limit \cite{xu2021}, while for the $\omega_1 = 0$ gravitational setting, the dephasing dynamics displays exponential decay with time as we shall see in the following.
The integral in the above expression can be evaluated analytically to give
\begin{align}
&\int_0^\infty d\omega \omega \coth \left(\frac{\beta \hbar \omega}{2} \right)\frac{\sin^2(\frac{\omega t}{2})}{\omega^2} e^{-{\omega}/{\omega_c}}\nn  \\
 =& \frac{1}{4}\ln \left(1+ t^2 \omega_c^2\right)\nn\\ +& \frac{1}{2}\ln \left[\frac{\Gamma^2 \left(\frac{1}{\beta \hbar \omega_c}+1 \right)}{\Gamma \left( \frac{1-it \omega_c}{\beta \hbar \omega_c}+1 \right)\Gamma \left(\frac{1+it \omega_c }{\beta \hbar \omega_c}+1 \right)}\right].
\end{align}
Taking the limit ${\beta \hbar \omega_c} \rightarrow \infty$ (i.e, upper cut-off frequency large compared to the bath temperature), we have
\begin{align}
\frac{\Gamma^2\left(\frac{1}{\beta \hbar \omega_c}+1 \right)}{\Gamma \left( \frac{1-it \omega_c }{\beta \hbar \omega_c}+1 \right)\Gamma \left(\frac{1+it \omega_c}{\beta \hbar \omega_c}+1 \right)} \rightarrow \frac{\beta \hbar }{\pi t} \sinh\left(\frac{\pi t}{\beta \hbar}\right).
\label{highcutoffapproxeq}
\end{align}
With approximation (\ref{highcutoffapproxeq}), Eq. (\ref{outerprod2eq}) becomes
\begin{align}
& |n(t)\rangle \langle n'(t)|=|n\rangle \langle n'| \exp\bigg[ -i\Omega(n-n')t+i\frac{C \Omega ^2}{\pi} \nn \\
&\times (n-n')(n+n'+1) \left[ \omega_c t - \tan^{-1}(\omega_c t) \right] -\frac{C\Omega^2}{\pi}\nn \\
&\times (n-n')^2 \left( \frac{1}{2}\ln \left(1+ t^2 \omega_c^2\right) +  \ln \left[\frac{\beta \hbar }{\pi t} \sinh\left(\frac{\pi t}{\beta \hbar}\right)\right]  \right)\bigg].
\label{outerprod3eq}
\end{align}

We now discuss the various terms appearing in Eq. (\ref{outerprod3eq}). First, note that the outer product is time-independent for $n=n'$, a consequence of the fact that the system oscillator Hamiltonian commutes with the system-bath interaction Hamiltonian. The first, pure imaginary term $-i\Omega (n-n')t$ in the argument of the exponential is just the free oscillator system evolution. The second pure imaginary term
\begin{align}
i\frac{C \Omega ^2}{\pi}(n-n')(n+n'+1) \left[ \omega_c t - \tan^{-1}(\omega_c t) \right]
\label{kerrtermeq}
\end{align}
is upper cut-off dependent and comprises both a linear term in system number, which renormalizes the system oscillator frequency $\Omega$, and a quadratic term in system number that is in  fact of the same form as the free evolution of a Kerr nonlinear oscillator expressed in the number state basis: 
\begin{align}
H=\hbar \Omega a^\dag a + \hbar\Lambda_{\mathrm{kerr}} (a^\dag a)^2.
\label{kerroscillator}
\end{align}
Thus, we should properly include a Kerr-type nonlinearity in our starting Hamiltonian (\ref{classical GR Hamiltonian}), with the environmentally induced term $i\frac{C \Omega ^2}{\pi}(n-n')(n+n')\omega_c t$ renormalizing the nonlinear interaction strength $\Lambda_{\mathrm{kerr}}$. The latter term may be thought of as somewhat analogous to the Newtonian gravitational self-interaction arising from the interaction of a matter system with its gravitating environment. Since we are primarily concerned with decoherence in the present work, we will neglect the quadratic in number term, supposing that it renormalizes an existing Kerr nonlinearity with resulting negligible renormalized coupling strength $\Lambda_{\mathrm{kerr}}$. For $t\gg\omega_c^{-1}$, the $\tan^{-1}(\omega_c t)$ term in (\ref{kerrtermeq}) tends to $\pi/2$; this term can be absorbed through a shift in the time coordinate: $t\rightarrow\tilde{t}=t-\pi/(2\omega_c)$.

Taking into account the system frequency and Kerr nonlinearity renormalizations as just described, Eq. (\ref{outerprod3eq}) simplifies to
\begin{align}
& |n(t)\rangle \langle n'(t)|=|n\rangle \langle n'| \exp\bigg( -i\Omega(n-n')t -\frac{C}{\pi} \nn \\
&\times (n-n')^2 \left( \frac{1}{2}\ln \left(1+ t^2 \omega_c^2\right) + \ln\left[ \frac{\beta \hbar }{\pi t} \sinh\left(\frac{\pi t}{\beta \hbar}\right)\right]  \right)\bigg),
\label{GRsolution}
\end{align}
where we have redefined the coupling constant as $C\rightarrow\tilde{C}=C \Omega^2$ and dropped the tilde. The real term on the second line of the argument of the exponential results in decoherence, i.e., exponential decay of the outer product for $n\neq n'$. In the high temperature (equivalently long time) limit corresponding to $t\gg \beta \hbar\gg \omega_c^{-1}$, Eq. (\ref{GRsolution}) can be approximated asymptotically as
\begin{align}
& |n(t)\rangle \langle n'(t)|=|n\rangle \langle n'| \exp\bigg( -i\Omega(n-n')t \nn \\
&-(n-n')^2 C\left[\frac{1}{\pi} \ln\left(\frac{\beta\hbar\omega_c}{2 \pi}\right)+(\beta\hbar)^{-1}t \right]\bigg).
\label{GRsolution2}
\end{align}
From Eq. (\ref{GRsolution2}), it is clear that the outer product terms for $n\neq n'$ decay exponentially with rate given by $(n-n')^2 C k_B T/\hbar$. Note however, that for early, `Planckian' (by analogy with gravity) times $t\lesssim \omega_c^{-1}$, the rate of decoherence is governed by the upper cut-off $\omega_c$, resulting in the logarithm term appearing in Eq. (\ref{GRsolution2}); depending on the magnitude of the ratio $\hbar\omega_c/{k_BT}\gg 1$, there may already be a significant `burst' of decoherence during the `Planckian' regime before the later, high temperature exponential decoherence regime. The fact that the decoherence rate depends on the upper cut-off frequency $\omega_c$ is a consequence of assuming an initial system-environment product state \cite{anglin1997}. The latter assumption is tantamount to supposing that the system initial state can be prepared  on time scales shorter than $\omega_c^{-1}$ (or equivalently, the system-environment interaction is switched on over a time scale shorter than $\omega_c^{-1}$).   While this may be possible for low energy, solid state system environments (i.e., phonons), for an actual gravitational environment with corresponding characteristic Planck time scale, the system state cannot be similarly isolated from the gravitational environment; an analysis which accounts for the system remaining correlated with the environment while its state is being prepared on timescales that are long compared with $\omega_c^{-1}$, is expected to result in a subsequent decoherence rate that does not depend on the upper cut-off frequency of the environment.  

\subsection{Decoherence}
\label{gravdecohsec}
In the following, we will use Eq. (\ref{GRsolution}) to determine the decoherence dynamics of the oscillator system for initial coherent state superpositions of the form
\begin{align}
\rho_{\mathrm{init}} = N\left( |\alpha_1 \rangle+| \alpha_{2} \rangle \right) \left(\langle \alpha_1 | + \langle \alpha_2 | \right) \otimes  \rho_{\mathrm{bath}},
\label{initcoheq}
\end{align}
where $|\alpha_1 \rangle$ and $|\alpha_2 \rangle$ denote coherent states and $N$ is the normalization constant. We consider coherent states since they describe most closely a cooled down, macroscopic oscillator  center of mass system. We emphasise that we do not rely on any of the approximations that are often invoked in the study of open quantum system dynamics (beyond assuming an initial product state). In particular, the following analysis is valid for both short/long time scales and high/low temperatures. 

\begin{figure}[htbp]
    \centering
\begin{subfigure}[b]{0.21\textwidth}
\centering
    \includegraphics[width=\textwidth]{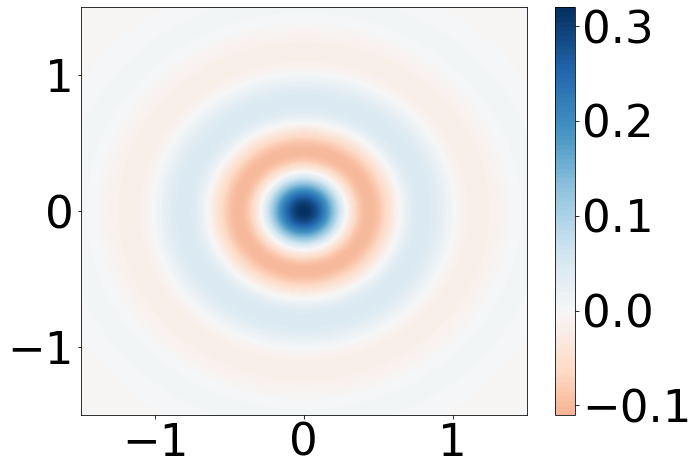}
    \caption{$\tau\rightarrow\infty$}
\end{subfigure}
\begin{subfigure}[b]{0.227\textwidth}
\centering
    \includegraphics[width=\textwidth]{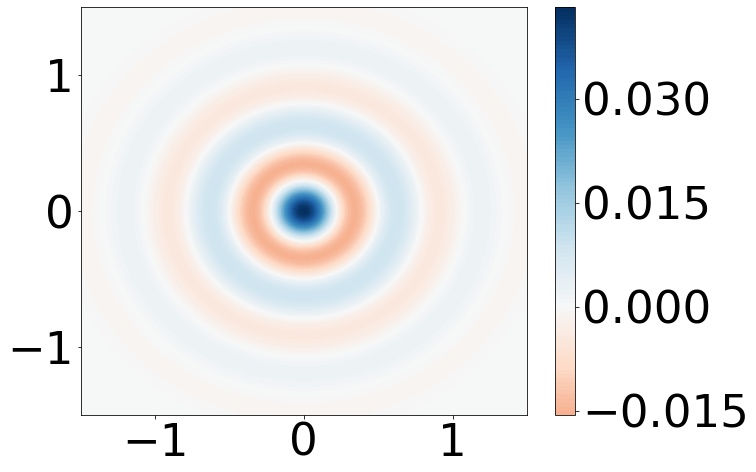}
    \caption{$\tau\rightarrow\infty$ }
\end{subfigure}
\caption{\label{ZoomGRwignerfig}Zoomed-in snapshots of the Wigner function: (a) $\alpha_1=-\alpha_2=3$; (b) $\alpha_1=3$, $\alpha_2=-5$.}
\end{figure}

Note from the form of (the exact) Eq.~(\ref{GRsolution}), 
that the system will evolve into a classical mixture of number states with probability coefficients that are identical to the coefficients of the initial system state. In contrast to other types of system-bath interaction where the final steady state of the system is usually temperature dependent, for the present model the temperature only determines how fast the system decoheres--not its long time limit steady state.
\begin{figure}[htbp]
    \centering
\graphicspath{{GRwigner/}}
\begin{subfigure}[b]{0.15\textwidth}
    \includegraphics[width=\textwidth]{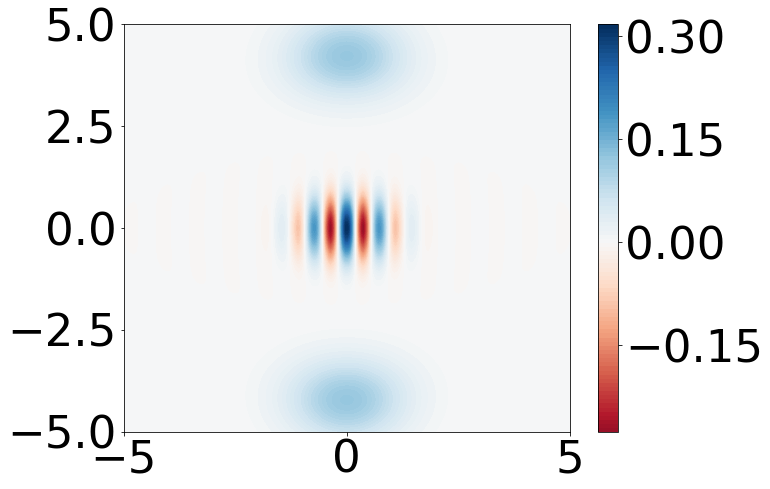}
    \caption{$\tau = \pi/2$}
\end{subfigure}
\hfill
\begin{subfigure}[b]{0.15\textwidth}

    \includegraphics[width=\textwidth]{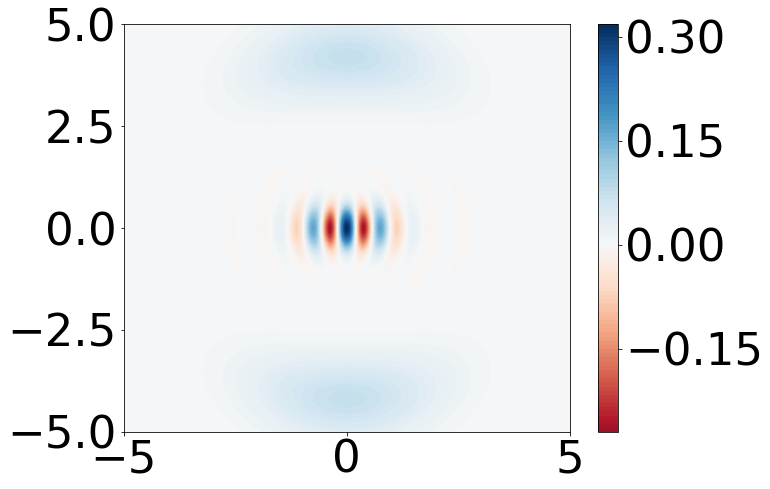}
    \caption{$\tau={9\pi}/{2}$ }
\end{subfigure}
\hfill
\begin{subfigure}[b]{0.144\textwidth}
    \includegraphics[width=\textwidth]{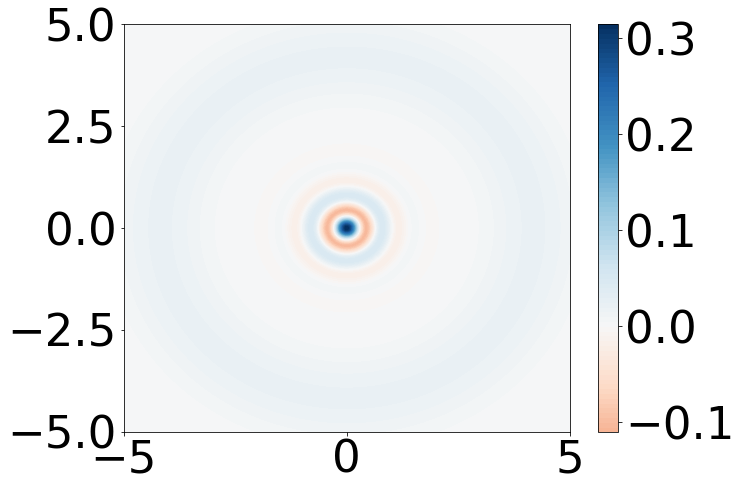}
    \caption{$\tau\rightarrow \infty$}
\end{subfigure}
\begin{subfigure}[b]{0.146\textwidth}

    \includegraphics[width=\textwidth]{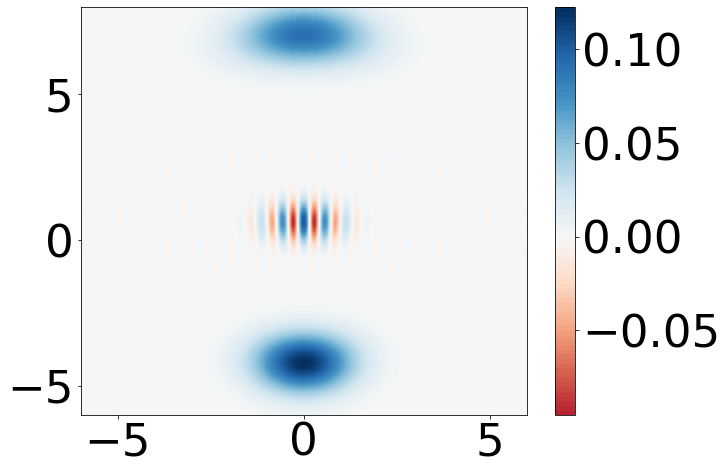}
    \caption{$\tau = \pi/2$}
\end{subfigure}
\hfill
\begin{subfigure}[b]{0.146\textwidth}

    \includegraphics[width=\textwidth]{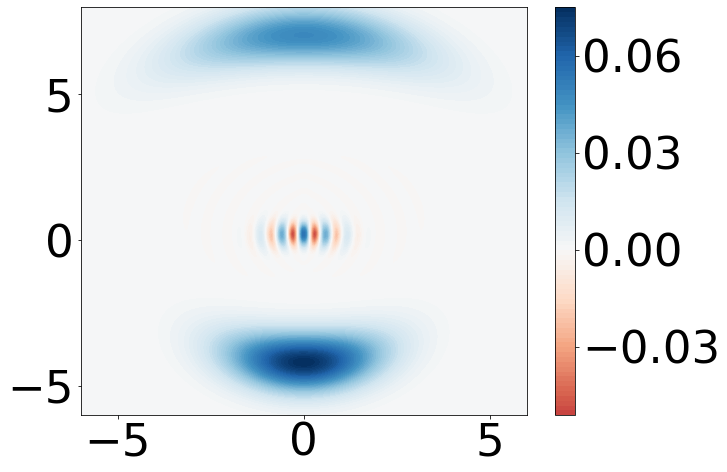}
    \caption{$\tau={9\pi}/{2}$}
\end{subfigure}
\hfill
\begin{subfigure}[b]{0.151\textwidth}
    \includegraphics[width=\textwidth]{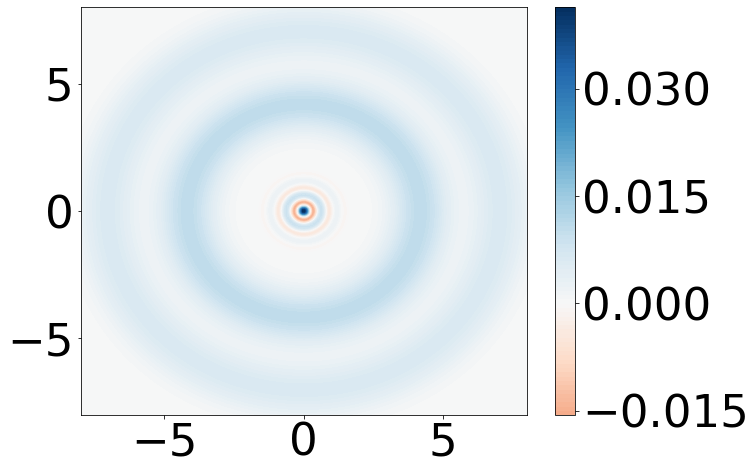}
    \caption{$\tau\rightarrow\infty$}
\end{subfigure}
\begin{subfigure}[b]{0.15\textwidth}
    \includegraphics[width=\textwidth]{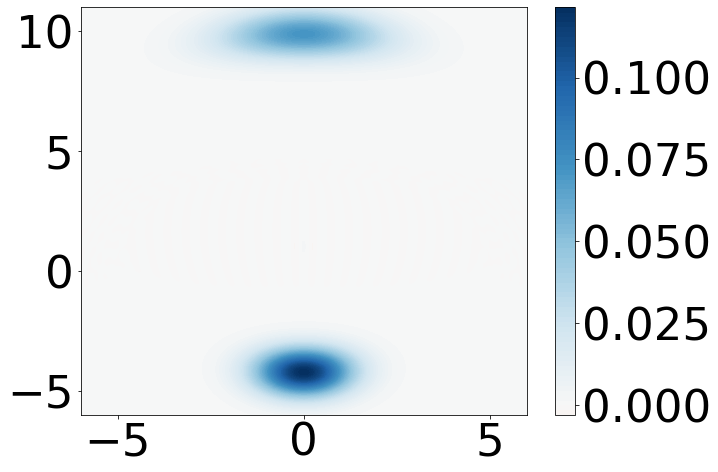}
    \caption{$\tau = \pi/2$}
\end{subfigure}
\begin{subfigure}[b]{0.162\textwidth}
    \includegraphics[width=\textwidth]{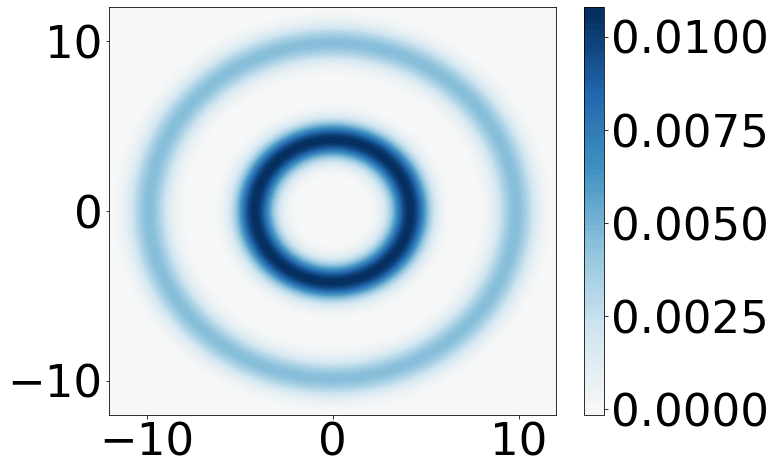}
    \caption{$\tau\rightarrow \infty$}
\end{subfigure}
\caption{\label{GRwignerfig} Wigner function snapshots at different times $\tau=\Omega t$ for the oscillator system. The horizontal coordinate is the dimensionless position $x\sqrt{M\Omega /\hbar}$ and the vertical coordinate is the dimensionless momentum $p/\sqrt{M\Omega \hbar}$. Example coherent state parameters are (a)-(c) $\alpha_1 = 3$, $\alpha_2 = -3$; (d)-(f) $\alpha_1 = 3$, $\alpha_2 = -5$; (g), (h) $\alpha_1 = 3$, $\alpha_2 = -7$. Other fixed system-bath parameters are: $\beta\hbar \Omega= 1$, ${\omega_c}/{\Omega} = 10^3$, $C/\pi= 0.001$.}
\end{figure}
From Eq. (\ref{GRsolution}), we also see that the decoherence rate is proportional to $(n-n')^2$ for a superposition of two number states $|n\rangle$ and $|n'\rangle$. Thus, for a superposition of coherent states, we expect that the larger the average energy difference, the more rapid the decoherence. This trend is apparent in the oscillator system Wigner function \cite{case2008} snapshots shown in Fig. \ref{GRwignerfig}. For the initial, example superposition state with $\alpha_1=3$, $\alpha_2=-7$, the negative Wigner function regions disappear in the long time limit (signifying loss of quantum coherence). On the other hand, for the initial example superposition states with nearby coherent state parameter magnitudes: $\alpha_1=-\alpha_2=3$,  $\alpha_1=3$ and $\alpha_2=-5$, negative Wigner function regions remain in the long time limit (signifying remaining quantum coherence), as is seen more clearly for the zoomed-in Fig. \ref{ZoomGRwignerfig}. Such trends are consistent with decoherence only resulting for initial spatial superpositions where the states making up the superposition have sufficiently distinct average energies; initial spatial superpositions with the same (or nearby) average energies for the states making up the superposition do not completely decohere.  
Note also from the Wigner function snapshots in Fig. \ref{GRwignerfig} and Fig. \ref{ZoomGRwignerfig} that the initial coherent superpositions phase-diffuse first into crescent-like regions and then eventually into rings.  This is consistent with the fact that, as mentioned above, the final state is always a mixture of number states.

The above findings are in accord with first investigations on the gravitational decoherence of massive scalar quantum field initial superposition states \cite{blencowe2013,anastopoulos2013}, where it is found that superpositions comprising distinct energy states decohere.

\begin{figure}[htbp]
    \centering
\begin{subfigure}[b]{0.22\textwidth}
\centering
\caption{}
    \includegraphics[width=\textwidth]{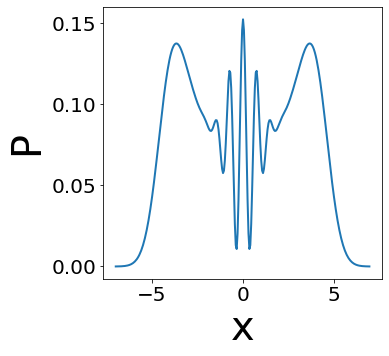}
\end{subfigure}
\hfill
\begin{subfigure}[b]{0.22\textwidth}
\centering
 \caption{}
    \includegraphics[width=\textwidth]{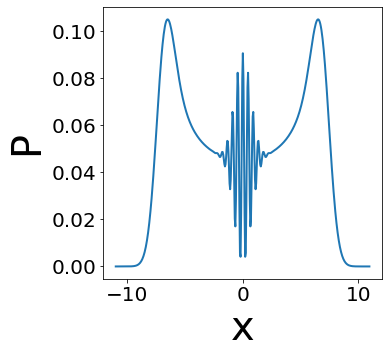}
\end{subfigure}

\begin{subfigure}[b]{0.22\textwidth}
\centering
 \caption{}
    \includegraphics[width=\textwidth]{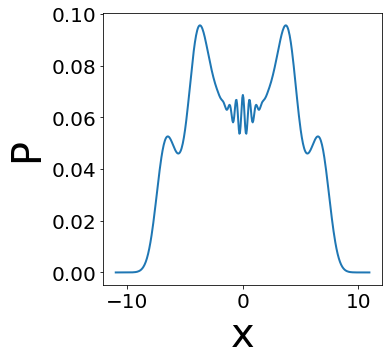}
\end{subfigure}
\hfill
\begin{subfigure}[b]{0.22\textwidth}
\centering
 \caption{}
    \includegraphics[width=\textwidth]{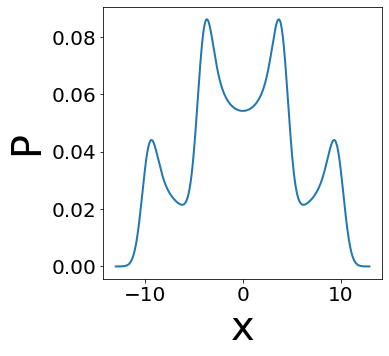}
\end{subfigure}
   \caption{\label{GRp_dis}Snapshots of the (unnormalized)  position probability density $P$ versus the dimensionless position coordinate $x\sqrt{M\Omega /\hbar}$ in the long time limit, steady state; the probability density is specified up to an overall normalization constant. (a) $\alpha_1=-\alpha_2 = 3$; (b) $\alpha_1 =-\alpha_2= -5$; (c) $\alpha_1 = 3$ and $\alpha_2  =-5$; (d) $\alpha_1 = 3$ and $\alpha_2 = -7$. The system-bath parameters are: $\beta\hbar \Omega= 1$, ${\omega_c}/{\Omega} = 10^3$, $C/\pi= 0.001$.}
\end{figure}

Following from the discussion in Sec. \ref{Introduction}, an operational way (i.e., in principle measurement procedure)  to quantify the coherence is through the system oscillator position detection probability density $P(x,t)=\langle x|\rho(t)|x\rangle$ [c.f. the full field-theoretic counterpart Eq. (\ref{detectoraveq})]  when the two (initially coherent) wavefunctions making up the superposition pass through each other at $x=0$; these time instants are $\tau_n=\Omega t_n={\pi}(n+1/2),\, n=0,1,2,\dots$ for the initial coherent state superposition examples considered above, as can be seen for the early time snapshots in Fig \ref{GRwignerfig}.  The presence of coherence is manifested in $P(x,t)$ having an oscillatory dependence about $x=0$. The latter operational approach corresponds to a two-slit inteference measurement, where the harmonic potential plays the role of the slits by (periodically) bringing the wavefunction components in the initial superposition together. Figure \ref{GRp_dis} shows the position probability distribution function in the long time limit, steady state for the various example initial coherent state superpositions; we can see that the probability density indicates interference fringes in the vicinity of $x=0$ consistent with the presence of negative-valued Wigner function regions shown in Fig. \ref{GRwignerfig}; the snapshots can be interpreted as the marginal probability distributions obtained by integrating over the momentum coordinate Wigner function distributions. In particular, the interference remains for $\alpha_1 \simeq -\alpha_2$, where the average energies of each coherent state making up the initial superposition are not too dissimilar. Note that the other, larger scale scale probability variations in Fig. \ref{GRp_dis}  are due to the final, steady state being a mixture of different number states, as mentioned earlier above.  

\begin{figure}[htbp]
    \centering
\begin{subfigure}[b]{0.23\textwidth}
\centering
\caption{}
    \includegraphics[width=\textwidth]{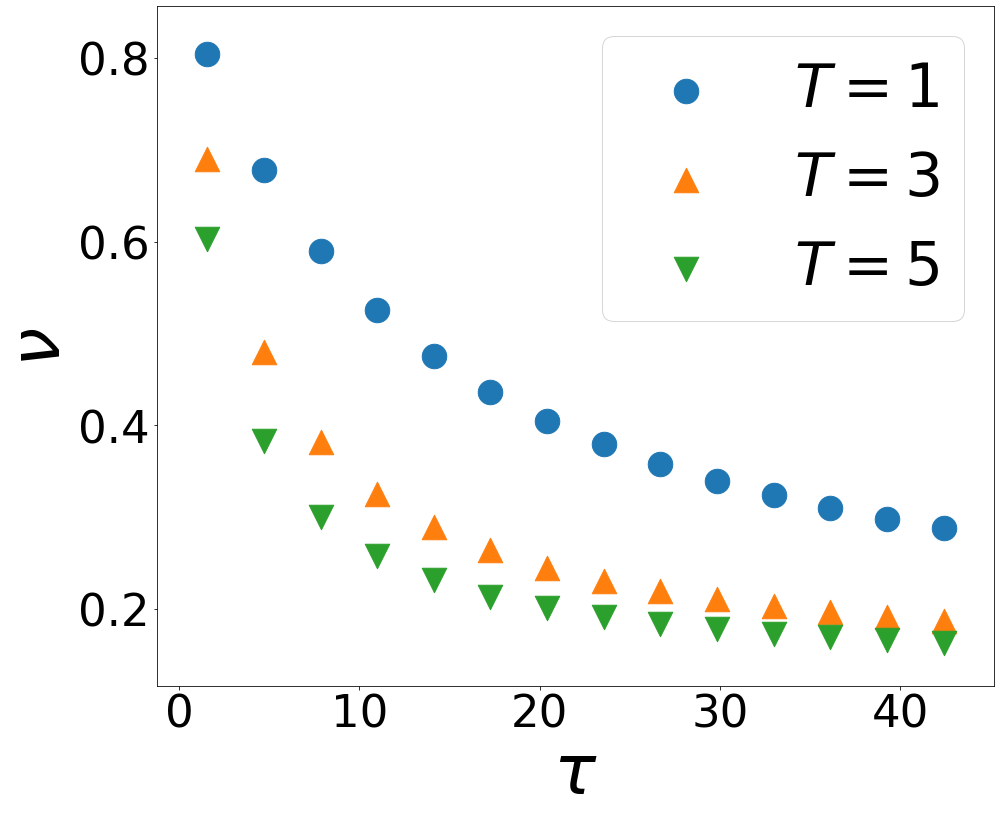}

\end{subfigure}
\vfill
\begin{subfigure}[b]{0.23\textwidth}
\centering
 \caption{}
    \includegraphics[width=\textwidth]{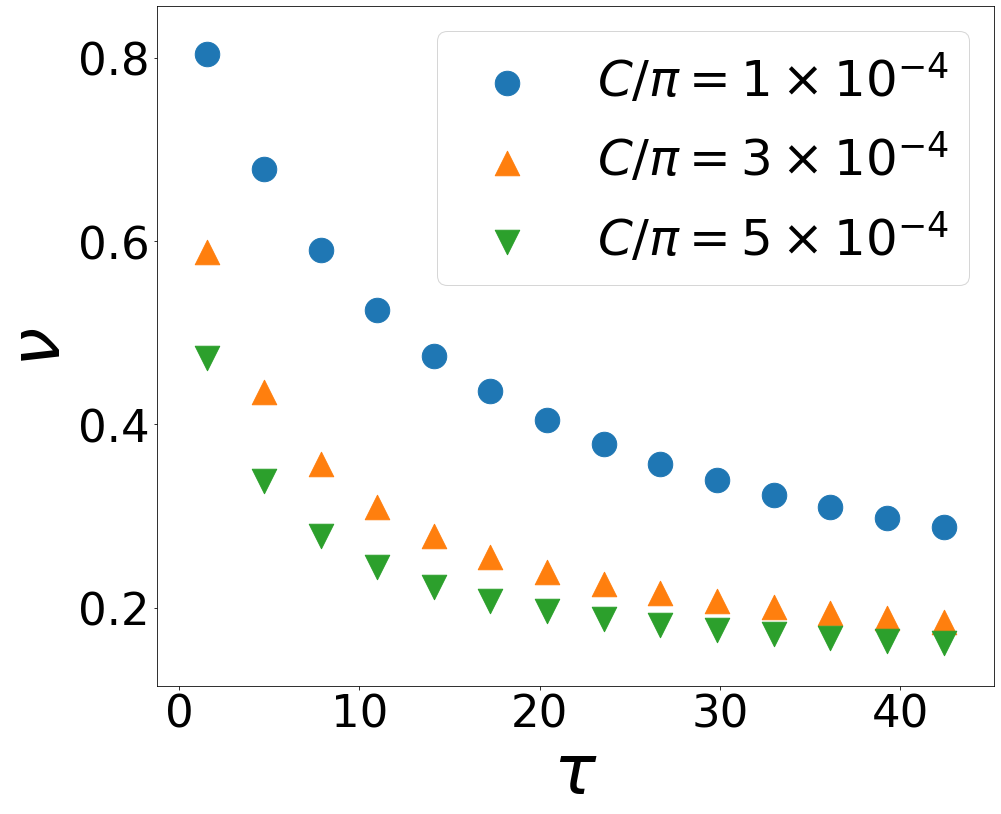}
   
\end{subfigure}
\begin{subfigure}[b]{0.23\textwidth}
\centering
 \caption{}
    \includegraphics[width=\textwidth]{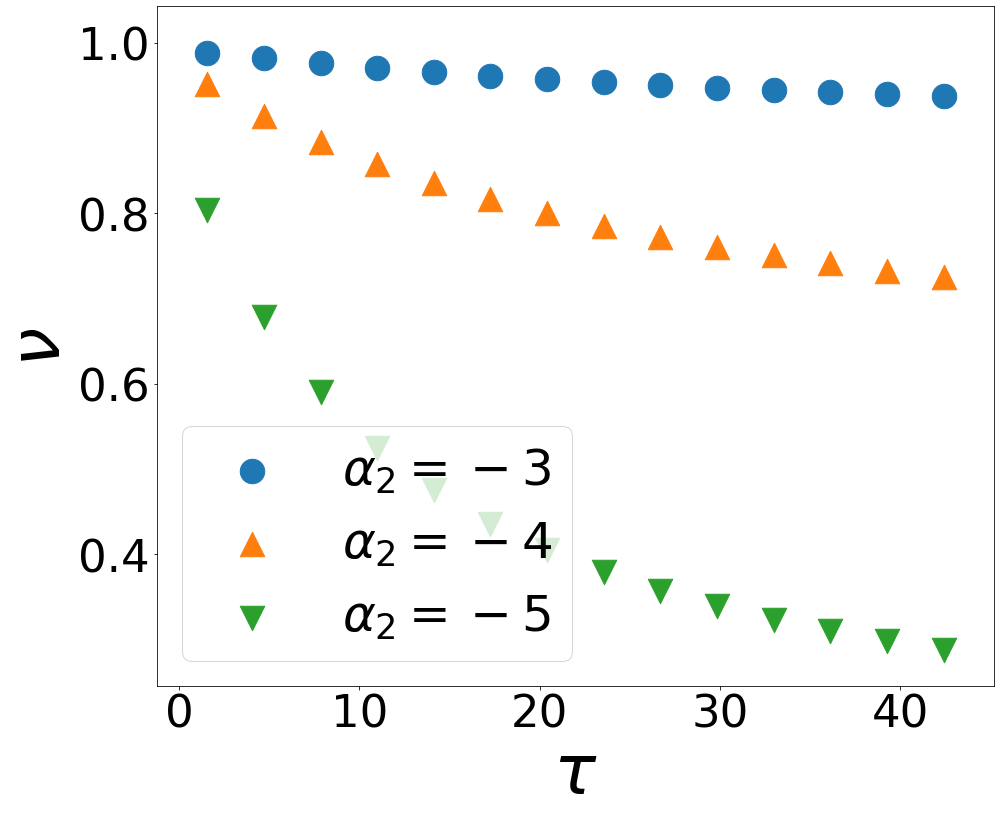}
 
\end{subfigure}
   \caption{\label{GRvisibility}Visibility as a function of dimensionless time $\tau=\Omega t$.
   The example parameters are (a) $\alpha_1 = 3$, $\alpha_2 = -5$, $C/\pi = 0.0001$, with dimensionless temperature $k_BT/(\hbar\Omega)$; (b)   $\alpha_1 = 3$, $\alpha_2 = -5$, $k_BT/(\hbar\Omega) = 1$; (c) $\alpha_1 = 3$, $C/\pi = 0.0001$, $k_BT/(\hbar\Omega) = 1$.}
\end{figure}

We adopt the commonly used `visibility' as a measure of the size of the interference fringes, defined as
\begin{align}
\nu = \frac{P_{{\mathrm{max}}} - P_{{\mathrm{min}}}}{P_{{\mathrm{max}}} + P_{{\mathrm{min}}}},
\label{visibilityeq}
\end{align}
where $P_{{\mathrm{max}}}$ is the central maximum of the probability density $P(x)$ at $x=0$ , and $P_{{\mathrm{min}}}$ is the first local minimum of the probability to the right (or left) of the central maximum. The decrease in visibility over time starting from the initial superposition state (\ref{initcoheq}), provides an operational, quantitative measure of decoherence; Fig. \ref{GRvisibility} gives the visibility as a function of time for various example, initial coherent state, bath temperature, and system-bath coupling parameters. As to be expected, the visibility decreases more rapidly the higher the temperature and the stronger the coupling. Also, the more dissimilar in magnitude $\alpha_2<0$ is from $\alpha_1>0$ (and hence the larger the average energy difference) in the initial coherent state superposition, the more rapid is the decrease in visibility. 

\section{Scalar QED model}
\label{qedsec}
In contrast to the scalar-gravity 0d toy model, the scalar QED toy model does not admit an exact, analytical solution for its quantum dynamics. We will therefore utilize various approximation methods towards solving for its quantum dynamics. In particular, we consider both quantum Langevin and master equation approaches, and approximations within these approaches that take advantage of the assumed weak system-bath interaction to show that the model maps onto the oscillator system with two-photon damping.

\subsection{Classical Langevin equation}

We start with the Lagrangian $L_{\mathrm{qed}}$ (\ref{0dqedeq}) and will first derive the Langevin equation  that describes the classical oscillator system dynamics interacting with the oscillator bath following the approach of Ref. \cite{cortes1985}. 
Expanding out the kinetic energy term of Eq. (\ref{0dqedeq}), we have
\begin{align}
L &= \frac{1}{2} M \dot x^2 - \frac{1}{2} M \Omega^2 x^2 + \sum_i \left(\frac{1}{2}m \dot q_i^2 - \frac{1}{2}m \omega_i^2 q_i^2\right) \nn \\ 
& + M x \dot x \sum_i \lambda_i q_i  + \frac{1}{2}M x^2 \sum_{i,j}\lambda_i \lambda_j q_i q_j,
\label{scalar QED L}
\end{align}
where we omit the `qed' subscript from now on. The system and bath momentum coordinates are
\begin{align}
&p = \frac{\partial L }{\partial \dot x} = M \dot x + M  x \sum_i \lambda_i q_i, \\
&p_i = \frac{\partial L}{\partial \dot q_i} = m \dot q_i,
\end{align}
and the model Hamiltonian is
\begin{align}
H  =&\frac{p^2}{2M} + \frac{1}{2} M\Omega^2 x^2 
+ \sum_i\left( \frac{p_i^2}{2 m} 
+ \frac{1}{2} m \omega_i^2 q_i^2\right) \nn \\ -&  x p \sum_i \lambda_i q_i.
\label{classical Hamiltonian}
\end{align}
Hamiltonian (\ref{classical Hamiltonian}) is to be compared with the gravity toy model Hamiltonian (\ref{gravham2eq}), which differs solely in the form of the system coordinate part of the interaction term; both models have in common a quadratic system coordinate coupling, to be contrasted with the usually studied oscillator system-oscillator bath model with interaction term that is linear in the coupled system and bath coordinates. 

Hamilton's equations for the system and bath coordinates are
\begin{align}
\dot p_i &= - m \omega_i^2 q_i + \lambda_i x p,\label{pieq}\\
\dot q_i &= \frac{p_i}{m},\label{qieq}\\
\dot p &= -M\Omega^2 x +  p \sum_i \lambda_i q_i,\label{peq}\\
\dot x &= \frac{p}{M} - x \sum_i \lambda_i q_i.\label{xeq}
\end{align}
Formally integrating the equations of motion (\ref{pieq}), (\ref{qieq}) for the bath coordinates  and expressing in terms of the system coordinates:
\begin{align}
q_i(t)&- \frac{\lambda_i x(t)p(t)}{m \omega_i^2}=\nn\\ 
 &\left[q_i(0) - \frac{\lambda_i}{m \omega_i^2}x(0)p(0) \right] \cos \omega_i t +\frac{p_i(0)}{m \omega_i} \sin \omega_i t\nn\\ 
&- \frac{\lambda_i}{m \omega_i^2} \int_0^t d\tau \cos \omega_i(t-\tau) \frac{d}{d\tau}\left( x(\tau)p(\tau)\right),
\label{bathsolneq}
\end{align}
where we have performed an integration by parts that allows to identify system renormalization and damping terms as we shall see below. Substituting the solution (\ref{bathsolneq}) for  $q_i(t)$ into the equations of motion (\ref{peq}), (\ref{xeq}) for the system coordinates leads to the following non-linear Langevin equations:
\begin{align}
\dot x &= \frac{\partial H^m}{\partial p} +  x \int_0^\tau d\tau K(t-\tau) \frac{d}{d\tau}\left(x(\tau)p(\tau) \right) -  x F(t),\label{langxeq} \\
\dot p &= -\frac{\partial H^m}{\partial x} -  p \int_0^t d\tau K(t-\tau) \frac{d}{d\tau} \left( x(\tau)p(\tau)\right)+ p F(t),\label{langpeq}
\end{align}
where the renormalized system Hamiltonian is
\begin{align}
H^m = \frac{p^2}{2M} +\frac{1}{2}M\Omega^2x^2 -  \sum_i \frac{\lambda_i^2}{2m \omega_i^2} x^2p^2,
\label{rensyshameq}
\end{align}
the bath memory kernel is 
\begin{align}
K(t-\tau) = \sum_i \frac{\lambda_i^2}{m \omega_i^2} \cos\omega_i (t-\tau),
\label{memoryeq}
\end{align}
and the bath random force function is 
\begin{align}
F(t) &= \sum_i \lambda_i\left(\left[q_i(0) - \frac{\lambda_i}{m \omega_i^2} x(0)p(0)\right]\cos \omega_i t\right. \nn \\
&\left.+\frac{p_i(0)}{m \omega_i}\sin \omega_i t \right).
\label{forcenoiseeq}
\end{align}
In particular, the first term on the right hand side of the equals sign in the Langevin equations  (\ref{langxeq}),(\ref{langpeq}) describes the Hamiltonian evolution, the second term describes nonlinear damping, and the third term the random force.  
Note that the bath induces a quartic anharmonic potential in the system Hamiltonian (\ref{rensyshameq}). Such a term is analogous to a Coulomb self-interaction potential in the scalar QED field system. After making the rotating wave approximation (RWA), the interaction term reduces to a Kerr-type nonlinearity [c.f. Eq. (\ref{kerroscillator})]. Together with `two-photon' damping [see Eq. (\ref{quantummastereq}) below], the resulting open system quantum dynamics can generate quantum states with negative-valued Wigner function regions in the long-time limit, starting from initial Gaussian states \cite{leghtas2015}. In the following, with decoherence dynamics our main subject of interest, we will neglect this induced potential energy term, supposing that it renormalizes an existing anharmonic potential with resulting negligible  renormalized coupling strength.  

Assuming a thermal equilibrium canonical ensemble distribution for the initial bath coordinates $q_i(0), p_i(0)$ in Eq. (\ref{forcenoiseeq}), it can be shown that the fluctuation dissipation relation (FDR) between the memory kernel and the random force  follows:
\begin{align}
\langle F(t) F(\tau)\rangle=k_B T K(t-\tau),
\label{fdreq}
\end{align}
where $k_B$ is Boltzmann's constant and $T$ is the bath temperature.
We shall assume that the bath responds rapidly on the time-scale of the system oscillator dynamics, so that memory kernel is approximated as 
$K(t-\tau) = \lambda^2 k_0 \delta(t-\tau)$, where $k_0$ is a constant. The Langevin equations (\ref{langxeq}), (\ref{langpeq})  then become 
\begin{align}
&\dot x = \frac{p}{M} + \frac{1}{2} \lambda^2 k_0 x\frac{d}{dt}(xp) - x F,\label{langx2eq}\\
&\dot p = -M\Omega^2 x  -\frac{1}{2} \lambda^2 k_0 p \frac{d}{dt}(xp) + p F,
\label{langp2eq}
\end{align}
with the FDR (\ref{fdreq}) taking the form
\begin{align}
\langle F(t)F(\tau) \rangle = k_B T k_0 \lambda^2 \delta(t - \tau).
\label{fdr2eq}
\end{align}

The above delta function-approximated memory kernel can be obtained from a   bath spectral density $n(\omega)$ with upper cut-off frequency $\omega_c$ in the limit $\omega_c\rightarrow\infty$. In particular, for a dense bath spectrum, we can approximate the sum over bath degrees of freedom  with a bath spectral frequency integral: 
\begin{equation}
\sum_i\lambda_i^2\left(\cdots\right)\rightarrow \int_0^{\infty} d\omega \lambda^2 n(\omega) \left(\cdots\right).
\end{equation}
Assuming a Lorentzian spectral density 
\begin{equation}
n(\omega)=  \frac{m k_0}{\pi} \frac{\omega^2 \omega_c^2}{\omega^2 +\omega_c^2},
\label{spectraldenseq}
\end{equation}
the memory kernel (\ref{memoryeq}) then becomes 
\begin{align}
K(t-\tau)=  \frac{\lambda^2 k_0\omega_c}{2} e^{-\omega_c|t-\tau|}.
\end{align}
Taking the infinite limit $\omega_c\rightarrow +\infty$, we obtain the above delta function-approximated memory kernel:
\begin{equation}
\lim_{\omega_c \rightarrow +\infty}K(t-\tau)=\lambda^2 k_0 \delta(t-\tau).
\end{equation}
Note that we could equally well have assumed a spectral density with exponential cut-off function instead, as for the gravity toy model [c.f. Eq. (\ref{bathspeceq})]; while the calculations are somewhat more straightforward for the Lorentzian spectral density, we do not expect any qualitative differences in the resulting system quantum dynamics. The motivation to use the Lorentzian spectral density here is purely calculational convenience.  
The classical, non-linear Langevin equations (\ref{langx2eq}), (\ref{langp2eq}) can be numerically solved as stochastic differential equations as we show in the following sections when comparing with the corresponding quantum dynamics.

%========================================
%========================================
\subsection{Quantum Langevin equation}
The quantum description is obtained through the correspondence principle where $x$, $p$ and $p_i$, $q_i$ become operators satisfying the canonical commutation relations:
\begin{equation}
    [x,p]=i\hbar,~  [x_i,p_j]=i\hbar\delta_{ij},
\end{equation}
with all other commutators vanishing. From Eq. (\ref{classical Hamiltonian}), the quantum Hamiltonian operator is 
\begin{align}
H =& \frac{p^2}{2M} + \frac{1}{2} M\Omega^2 x^2 + \sum_i \left(\frac{p_i^2}{2 m}+\frac{1}{2} m \omega_i^2 q_i^2\right) \nn \\
&- \frac{1}{2} \sum_i\lambda_i q_i (x p+p x),
\label{quantum Hamiltonian}
\end{align}
where the interaction term on the second line is symmetrized in $x$ and $p$ in order that $H$ is Hermitian. 
Formally integrating Heisenberg's equations of motion for  the bath operators, we obtain the following quantum Langevin equations for the system position and momentum operators:
\begin{align}
\dot x &= \frac{p}{M} -  \sum_i  \frac{\lambda_i^2}{2m \omega_i^2}x\left(xp +px\right) \nn \\
&+ \frac{1}{2} x \int_0^\tau d\tau K(t-\tau)\frac{d}{d\tau}\left(x p + p x \right)  -  F(t)x,
\label{xquantumlangeq}
\end{align}
\begin{align}
\dot p &= -M \Omega^2 x +  \sum_i \frac{\lambda_i^2}{2m \omega_i^2}p \left(xp +px\right) \nn \\
&- \frac{1}{2} p  \int_0^\tau d\tau K(t-\tau)\frac{d}{d\tau}\left(x p + p x \right) +  F(t)p,
\label{pquantumlangeq}
\end{align}
where the force noise operator is given by Eq. (\ref{forcenoiseeq}) with the system/bath coordinates and momenta replaced by their corresponding operators, and $K(t-\tau)$ is the memory kernel given by Eq.~(\ref{memoryeq}).
It is convenient to express the quantum Langevin equations in terms of the system creation and annihilation operators which are defined through the usual relations
$x = \sqrt{\frac{\hbar}{2M\Omega}} (a + a^{\dag})$,  $p = i \sqrt{\frac{M\Omega\hbar }{2}}(a^\dag -a)$:
\begin{align}
\dot a &= - i \Omega a - \frac{i \hbar  }{2}\sum_i \frac{\lambda_i^2}{m \omega_i^2} a^\dag\left(a^{\dag 2} -a^2\right) -  F(t)a^\dag  \nn \\
&+ \frac{i \hbar }{2}a^\dag \int_0^t d\tau K(t-\tau) \frac{d}{d \tau}\left(a^{\dag 2} - a^2\right).
\label{full quantum langevin eq}
\end{align}

Under conditions of weak system-environment coupling, Eq. (\ref{full quantum langevin eq}) can be simplified by applying the RWA as we now show. Making the substitution $a(t) = A(t)e^{-i\Omega t}$ in Eq. (\ref{full quantum langevin eq}),
we obtain
\begin{align}
\dot A &= - \frac{i \hbar  }{2}\sum_i\frac{\lambda_i^2}{m \omega_i^2}A^{\dag}\left(e^{4i\Omega t}A^{\dag 2}  - A^2 \right)  \nn \\
&+  \frac{i\hbar}{2}A^{\dag} e^{2i\Omega t}  \int_0^t d\tau K(t-\tau) \frac{d}{d\tau}\left(A^{\dag 2} e^{2i\Omega \tau}\right. \nn \\
&\left.- A^2 e^{-2i\Omega \tau} \right) - e^{2i\Omega t} F(t)A^{\dag}.
\label{rotating_lang_eq}
\end{align}
Dropping fast rotating terms, neglecting time derivatives of $A(\tau)$ (since $A$ evolves at much slower rates than  $\Omega$), and setting $A(\tau)=A(t)$ (Markov approximation),  Eq.~(\ref{rotating_lang_eq}) becomes approximately
\begin{align}
\dot A &= \frac{i\hbar}{2} \sum_i\frac{\lambda_i^2}{m \omega_i^2}A^{\dag} A^2  -  e^{2i\Omega t}F(t) A^{\dag}   \nn \\
&- {\hbar\Omega} \int_0^t d\tau K(t-\tau)e^{2i\Omega(t-\tau)}A^{\dag}(t) A(t)^2.
\label{qlangeq}
\end{align}
Utilizing the Lorentzian spectral density (\ref{spectraldenseq}) [with Eq.~(49)], Eq. (\ref{qlangeq}) becomes
\begin{equation}
\dot A = \frac{i\gamma\omega_c }{2 \Omega} A^\dag A^2   - \frac{\gamma \omega_c}{\omega_c-2i\Omega} A^{\dag} A^2  
 -  e^{2i\Omega t} F(t) A^\dag,
 \label{qlang2eq}
\end{equation}
where we have dropped fast oscillating terms and where $\gamma=\hbar\Omega\lambda^2 k_0/2$. 
For $\omega_c\gg\Omega$ and neglecting the anharmonic interaction term,  Eq. (\ref{qlang2eq}) simplifies to
\begin{equation}
\dot A = -\gamma A^\dag A^2  - e^{2i\Omega t}F(t) A^\dag.
\label{qlang3eq}
\end{equation}
Defining the noise operator as 
\begin{equation}
b(t) = \frac{- e^{2i\Omega t} F(t)} {2\sqrt{\gamma}}
\label{noiseopeq}
\end{equation}
and utilizing Eqs. (\ref{forcenoiseeq}), (\ref{spectraldenseq}) and the RWA, the usual noise operator (anti)commutation rules follow:
\begin{equation}
\begin{split}
[b(t),b^{\dag}(t')] &= \delta(t-t'),\\
\{b(t), b^{\dag}(t')\} &= \delta(t-t')\big[2n(2\Omega)+1 \big],
\end{split}
\end{equation}
 where the Bose-Einstein thermal average occupation number is evaluated at twice the system oscillator frequency: $n(2\Omega) = \left(e^{2\hbar \Omega/k_B T} -1\right)^{-1}$. Finally, transforming back to the non-rotating frame, $A(t) = a(t)e^{i\Omega t}$, we obtain our desired, RWA quantum Langevin equation:
\begin{equation}
\dot a  =  -i\Omega a -\gamma a^\dag a^2 + 2\sqrt{\gamma} e^{-2i\Omega t} b a^\dag.
\label{RWA quantum langevin eq}
\end{equation}
From Eq. (\ref{RWA quantum langevin eq}), we see that the parameter $\gamma$ has the dimensions of inverse time and characterizes the strength of a nonlinear damping term, while the third term is the nonlinear force noise operator.  Equation (\ref{RWA quantum langevin eq}) can be solved numerically as a quantum stochastic differential equation or approximately by first deriving the equations for the various moments in $a$, $b$, and their Hermitian conjugates and truncating at some order.

\subsection{Quantum master equation}
An alternative way to express the quantum dynamics is via the quantum master equation, where the time evolution is given by the oscillator system reduced density matrix. 
To second order in the interaction potential and assuming that the bath responds much more rapidly than the system oscillation timescale   (Born-Markov approximation), the master equation for system density matrix $\rho$ in the interaction picture is approximately \cite{carmichael1999,petruccione2002,gardiner2004},
\begin{align}
\frac{d\rho}{dt} = -\frac{1}{\hbar^2} \int _0^t dt' {\mathrm{Tr}}_B\left [V(t),[V(t'),  \rho(t)\otimes \rho_B] \right],
\label{original master eq}
\end{align}
where $\rho_B$ is the initial thermal state of the bath, ${\mathrm{Tr}}_B$ denotes the trace over the bath state and $V(t)$ is the system-bath interaction Hamiltonian expressed in the interaction picture:
\begin{align}
V(t) =& \frac{i\hbar }{2}\sum_i \lambda_i\sqrt{\frac{\hbar}{2m \omega_i}}e^{iH_0 t} \left(b_i^\dag +b_i\right)\left(a^{\dag 2} -a^2\right) e^{-iH_0 t} \nn \\
  =&\frac{i\hbar \lambda }{2}\sum_i\lambda_i\sqrt{\frac{\hbar}{2m_i \omega_i}}\left(b_i^\dag e^{i\omega_i t} +b_ie^{-i\omega_i t}\right)\nn\\ &\times\left(a^{\dag 2}  e^{2i\Omega t} - a^2 e^{-2i\Omega t} \right),
\end{align}
where we have introduced creation/annihilation operators for the bath: $q_i = \sqrt{\frac{\hbar}{2m\omega_i}} (b_i + b_i^{\dag})$, $p_i = i \sqrt{\frac{m\omega_i\hbar }{2}}(b_i^\dag -b_i)$. In order to simplify the next steps, we introduce the following shorthand notation:
\begin{align}
{\mathcal{A}}(t) &= a^{\dag 2}  e^{2i\Omega t} - a^2 e^{-2i\Omega t}\label{aeq} \\
{\mathcal{B}}(t) & = \sum_i\lambda_i\sqrt{\frac{\hbar}{2m \omega_i}}\left(b_i^\dag e^{i\omega_i t}+b_ie^{-i\omega_i t} \right).
\label{beq}
\end{align}
Expanding out Eq. (\ref{original master eq}) and substituting in Eqs. (\ref{aeq}) and (\ref{beq}),  we obtain:
\begin{align}
&\frac{d \rho}{dt} = \frac{1}{4} \int _0^t dt' \left\{ \left[{\mathcal{A}}(t){\mathcal{A}}(t') \rho - {\mathcal{A}}(t')\rho {\mathcal{A}}(t) \right]\langle {\mathcal{B}}(t){\mathcal{B}}(t')\rangle\right.\nn\\
&+ \left.\left[\rho {\mathcal{A}}(t'){\mathcal{A}}(t) - {\mathcal{A}}(t)\rho {\mathcal{A}}(t') \right] \langle {\mathcal{B}}(t'){\mathcal{B}}(t) \rangle \right\},
\label{mastereq2}
\end{align}
where 
\begin{align}
\langle {\mathcal{B}}(t){\mathcal{B}}(t') \rangle =& \sum_i \frac{\lambda_i^2\hbar}{2m \omega_i} \left[ (n(\omega_i) + 1)e^{-\omega_i (t-t')}\right. \nn \\
 &\left.+ n(\omega_i) e^{i \omega_i (t-t')} \right]. 
\end{align}
Using the bath spectral density (\ref{spectraldenseq}) [with Eq.~(49)] and applying the RWA, we obtain the following quantum master equation:
\begin{align}
\frac{d \rho}{dt} =& i\Omega[\rho,a^\dag a] + \frac{\gamma}{2}(n+1)\left([a^2\rho, a^{\dag 2}] + [a^2,\rho a^{\dag 2}]\right) \nn \\
&+\frac{\gamma}{2}n \left([a^{\dag 2}\rho,a^2]+ [a^{\dag 2},\rho a^2] \right),
\label{quantummastereq}
\end{align}
where $n=n(2\Omega)=\left(e^{2\hbar \Omega/k_B T} -1\right)^{-1}$. In Eq. (\ref{quantummastereq}), we recognize an oscillator subject to `two-photon' damping.

As a consistency check, we can obtain an equation for the expectation value of $a$ starting either from the quantum Langevin equation (\ref{RWA quantum langevin eq}) with $\langle a\rangle={\mathrm{Tr}}\left(a(t)\rho(0)\right)$ or from the master equation (\ref{quantummastereq}) with $\langle a\rangle={\mathrm{Tr}}\left(a(0)\rho(t)\right)$; both approaches coincide to give
\begin{align}
\langle  \dot a \rangle = -i \Omega \langle a \rangle -\gamma \langle a^\dag a^2 \rangle + 2\gamma n(2\Omega)\langle a\rangle.
\label{avaeq}
\end{align}

\subsection{Validity of the RWA and quantum vs classical dynamics}
Starting with the 0d analogue scalar QED model Lagrangian (\ref{0dqedeq}), in the previous sections we derived a Markov approximated classical Langevin equation (\ref{langx2eq}), (\ref{langp2eq}), a Markov-RWA quantum  Langevin equation (\ref{RWA quantum langevin eq}), and a corresponding Markov-RWA quantum master equation (\ref{quantummastereq}). In the following, we will test the validity of the RWA at the classical level, as well as compare the classical versus RWA quantum dynamics for the averaged quantities $\langle a\rangle$ and $\langle a^{\dag} a\rangle$.

\begin{figure}[htbp]
    \centering
\begin{subfigure}[b]{0.23\textwidth}
\centering
\caption{}
    \includegraphics[width=\textwidth]{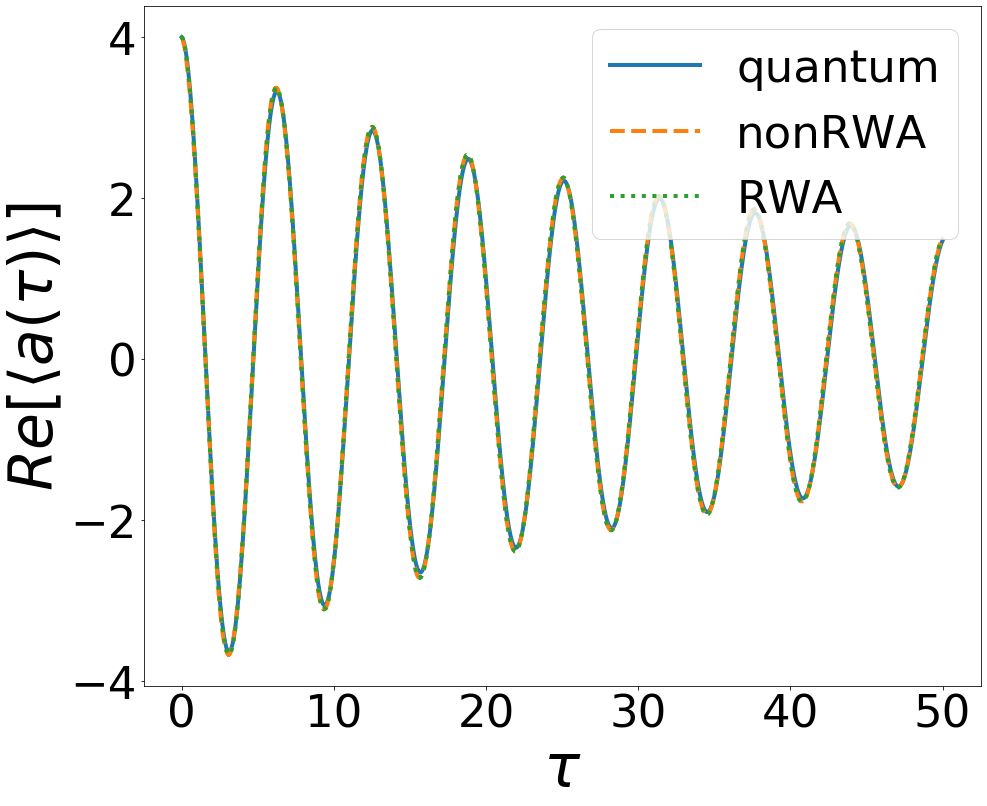}
\end{subfigure}
\begin{subfigure}[b]{0.23\textwidth}
\centering
\caption{}
    \includegraphics[width=\textwidth]{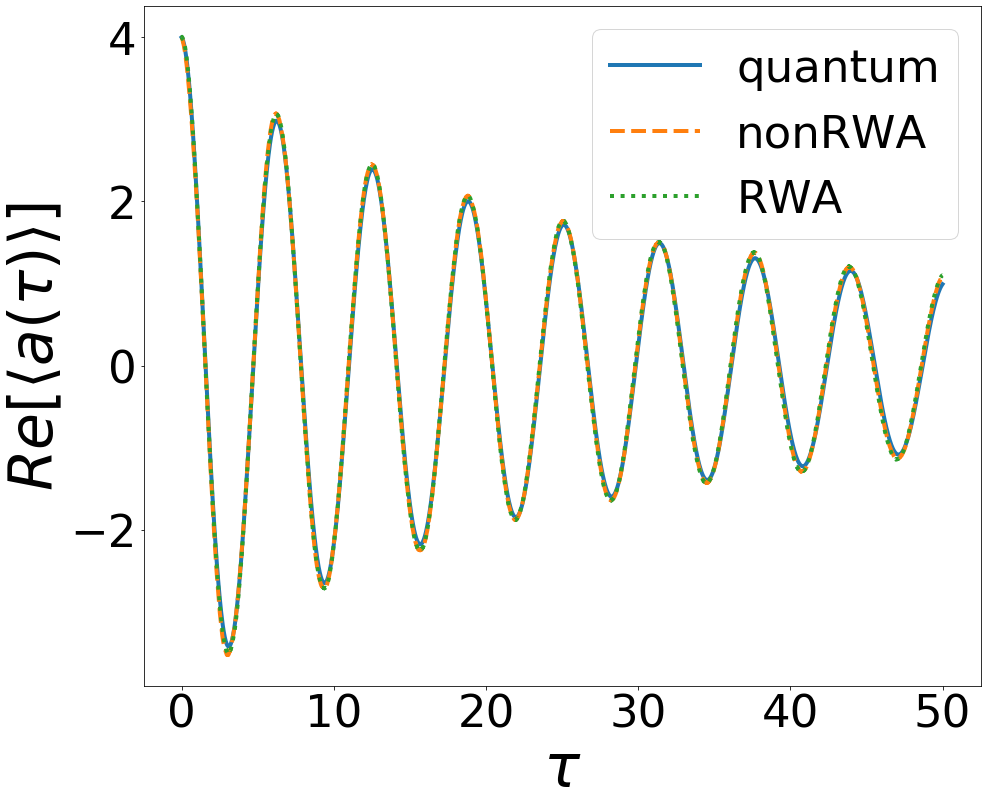}
    
\end{subfigure}
\vfill
\begin{subfigure}[b]{0.23\textwidth}
\centering
 \caption{}
    \includegraphics[width=\textwidth]{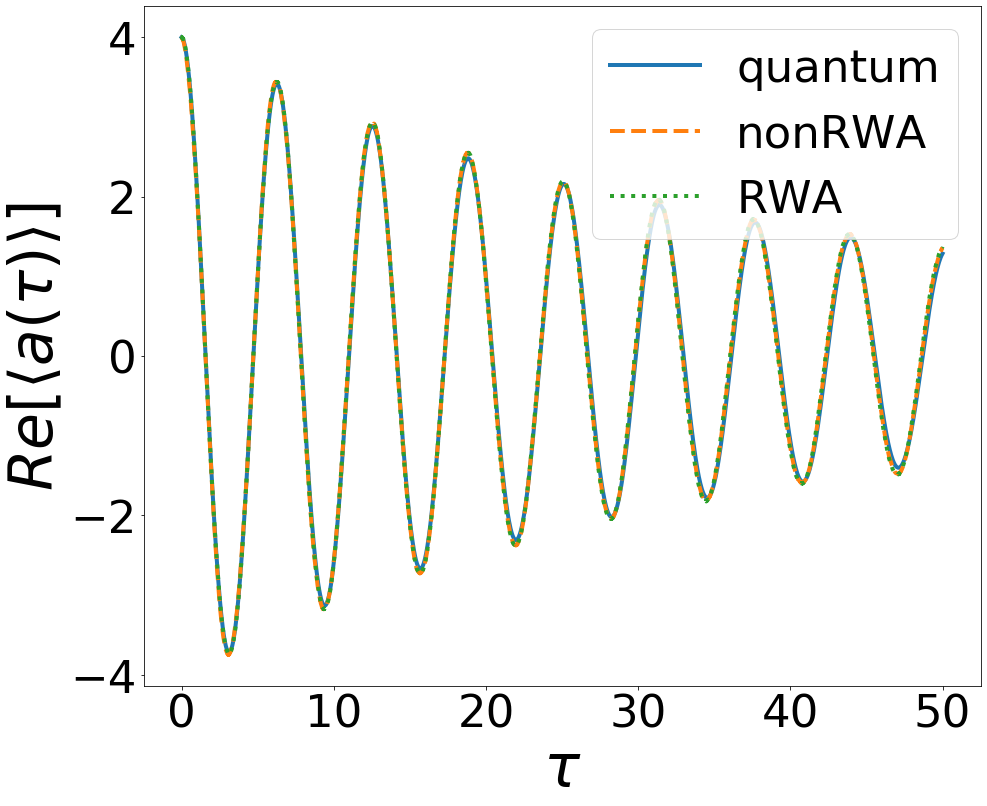}
   
\end{subfigure}
\begin{subfigure}[b]{0.23\textwidth}
\centering
 \caption{}
    \includegraphics[width=\textwidth]{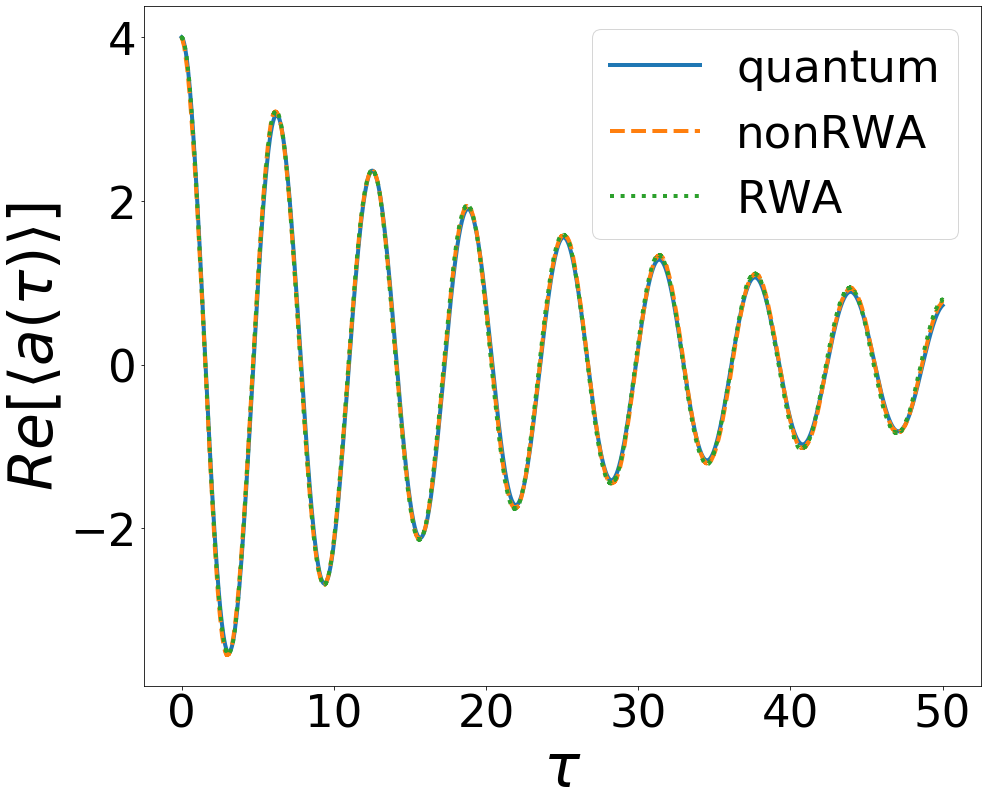}
   
\end{subfigure}
   \caption{\label{amplfig}Plots of the dimensionless average position $\langle x\rangle \sqrt{M\Omega/(2\hbar)}={\mathrm{Re}}\left[\langle a\rangle\right]$ as a function of dimensionless time $\tau=\Omega t$. The initial value $\alpha=a(0)=4$ in each plot. The time evolution of the classical amplitude $a(t)$ is the result of  averaging over 3000 stochastic trajectories.  The example parameters are (a) $Q^{-1}$ = 0.003, $n = 3$; (b) $Q^{-1}$ = 0.005, $n = 3$; (c) $Q^{-1}$ = 0.003, $n = 5$;  (d) $Q^{-1}$ = 0.005, $n = 5$. 
   }
\end{figure}

\begin{figure}[htbp]
    \centering
\begin{subfigure}[b]{0.23\textwidth}
\centering
  \caption{}
    \includegraphics[width=\textwidth]{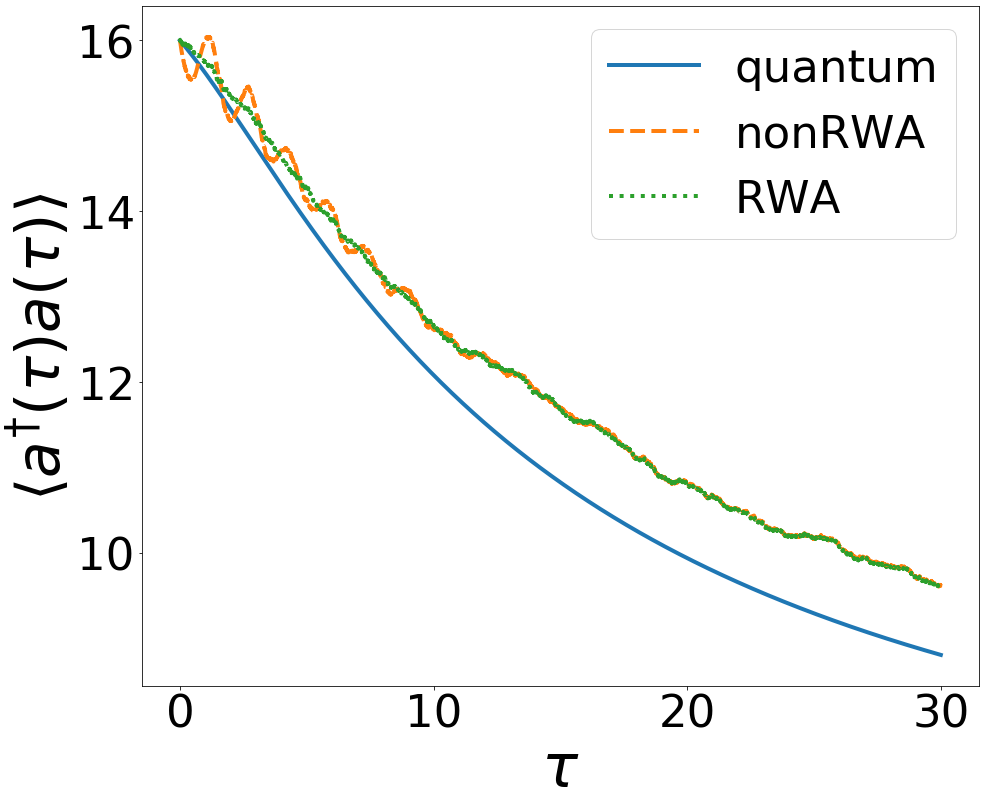}
  
\end{subfigure}
\begin{subfigure}[b]{0.23\textwidth}
\centering
 \caption{}
    \includegraphics[width=\textwidth]{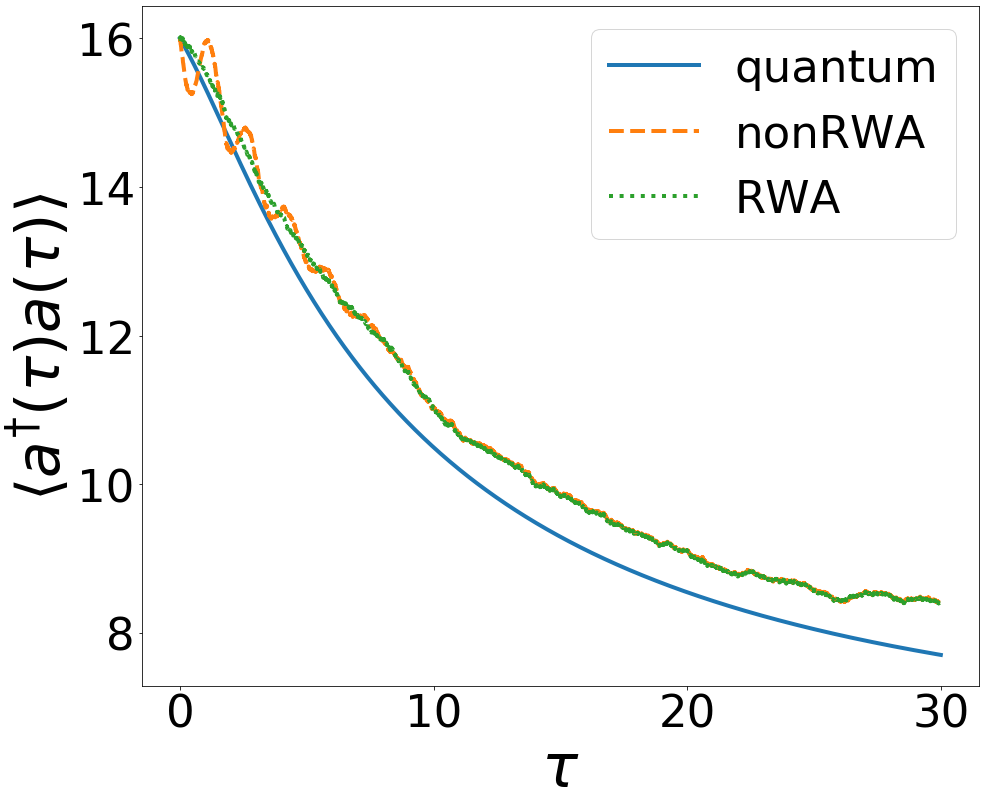}
   
\end{subfigure}
\vfill
\begin{subfigure}[b]{0.23\textwidth}
\centering
 \caption{}
    \includegraphics[width=\textwidth]{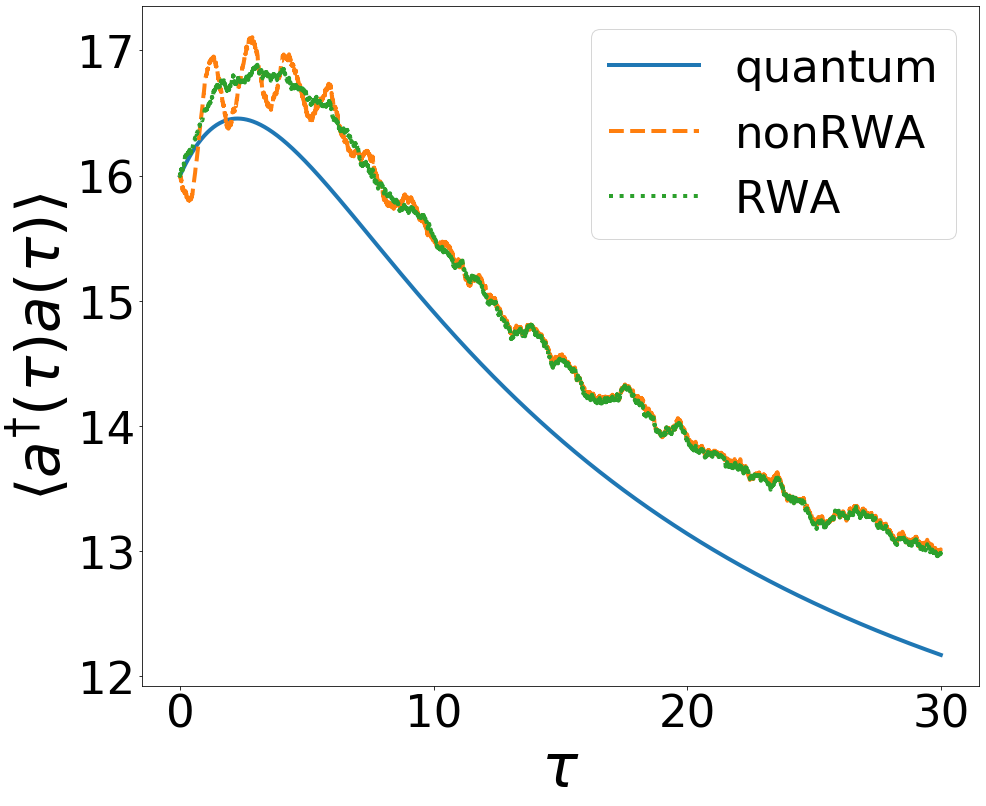}
   
\end{subfigure}
\begin{subfigure}[b]{0.23\textwidth}
\centering
 \caption{}
    \includegraphics[width=\textwidth]{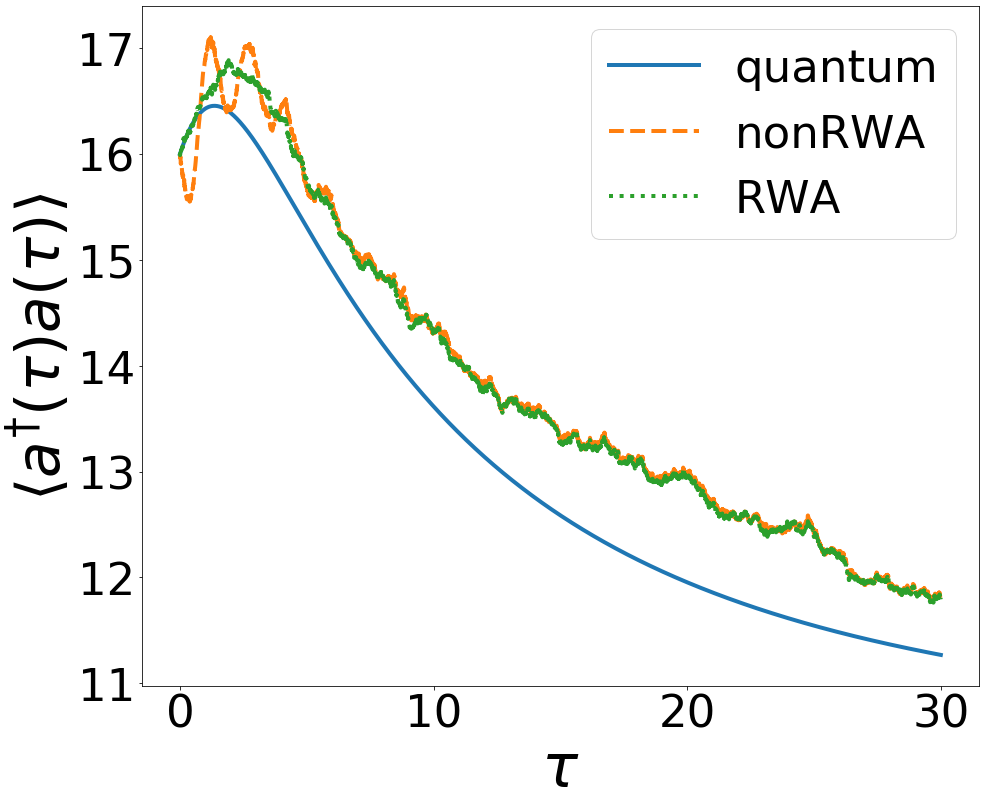}
   
\end{subfigure}
 \caption{\label{numfig}Plots of the average system number $\langle a^{\dag} a\rangle$ as a function of dimensionless time $\tau=\Omega t$. The initial value $\alpha=a(0)=4$ in each plot. The time evolution of the classical absolute amplitude squared $a(t)a^*(t)$ is the result of  averaging 5000 stochastic trajectories.   The example parameters are (a) $Q^{-1} = 0.003$, $n = 3$; (b) $Q^{-1} = 0.005$, $n = 3$; (c) $Q^{-1} = 0.003$, $n = 5$;  (d) $Q^{-1} = 0.005$, $n = 5$. }
\end{figure}

It is convenient to express the classical Langevin equations (\ref{langx2eq}), (\ref{langp2eq}) in terms of the complex coordinates $(a,a^*)$ corresponding to the quantum annihilation/creation operators:
\begin{equation}
\dot a = -i\Omega a + \frac{i\gamma}{2\Omega} \frac{d}{dt} (a^* a^* - a a) a^* -\sqrt{\frac{2\gamma}{\hbar \Omega}}\tilde F a^*,
\label{classalangeq}
\end{equation}
where $\tilde F = \frac{F}{\lambda\sqrt{k_0}}$, so that $\langle \tilde F(t) \tilde F(\tau)\rangle = k_B T \delta(t -\tau)$. The corresponding classical RWA Langevin equation is [c.f. Eq. (\ref{RWA quantum langevin eq})]:
\begin{equation}
\dot a = -i\Omega a -\gamma a^2 a^*  -\sqrt{\frac{2\gamma}{\hbar \Omega}}\tilde F a^*.
\label{classarwalangeq}
\end{equation}
In order to solve the non-RWA (\ref{classalangeq})  and RWA (\ref{classarwalangeq}) Langevin equations, we treat them as classical stochastic differential equations:
\begin{equation}
da = -i\Omega a dt + \frac{i\gamma}{2\Omega} \frac{d}{dt}(a^* a^*- aa)a^* dt -\sqrt{\frac{2 \gamma k_B T}{\hbar \Omega}}a^*dW,
\label{stocheq}
\end{equation}
\begin{equation}
da =-i \Omega a dt -\gamma a^2a^* dt -\sqrt{\frac{2 \gamma k_B T}{\hbar \Omega}}a^*dW,
\label{stochrwaeq}
\end{equation}
where $W$ is the standard Wiener process, i.e., a continuous-time random walk  \cite{gardiner1985}. Figures \ref{amplfig} and \ref{numfig} give numerical solutions to these classical stochastic equations as well as to the quantum master equation (\ref{quantummastereq}) (the latter solved using QuTiP \cite{johansson2013qutip}) for a range of damping parameters $\gamma$ and bath temperatures $T$. These parameters are respectively expressed in terms of the dimensionless $Q=\Omega/\gamma$ factor and thermal average bath occupation number $n$. The quantum system is initially in a coherent state $|\alpha \rangle$ for which $a |\alpha \rangle = \alpha |\alpha \rangle$, while the corresponding classical system is given an initial amplitude $a(0)=\alpha$, in order to allow a direct comparison between the quantum and classical dynamics.
From Fig. \ref{amplfig}, it can be seen that increasing $n$ and $Q^{-1}$ both lead to faster decay of the amplitude, signalling the non-linear nature of the damping and noise terms in the system Langevin and quantum master equations. It can also be seen that the difference between non-RWA, RWA and classical vs quantum is barely visible with the chosen parameters. However, such differences clearly show up in Fig. \ref{numfig} where we consider the time evolution of the average system number $\langle a^{\dag}a \rangle$. In particular, throwing away fast rotating terms due to the RWA results in smoothing of the oscillating behaviour of the non-RWA time evolution of $\langle a^{\dag}a \rangle$. Furthermore, the quantum simulation of $\langle a^{\dag}a \rangle$ decays faster than the corresponding classical approximation.

\subsection{Decoherence}
\begin{figure}[htbp]
    \centering
\begin{subfigure}[t]{0.15\textwidth}
\centering
    \includegraphics[width=\textwidth]{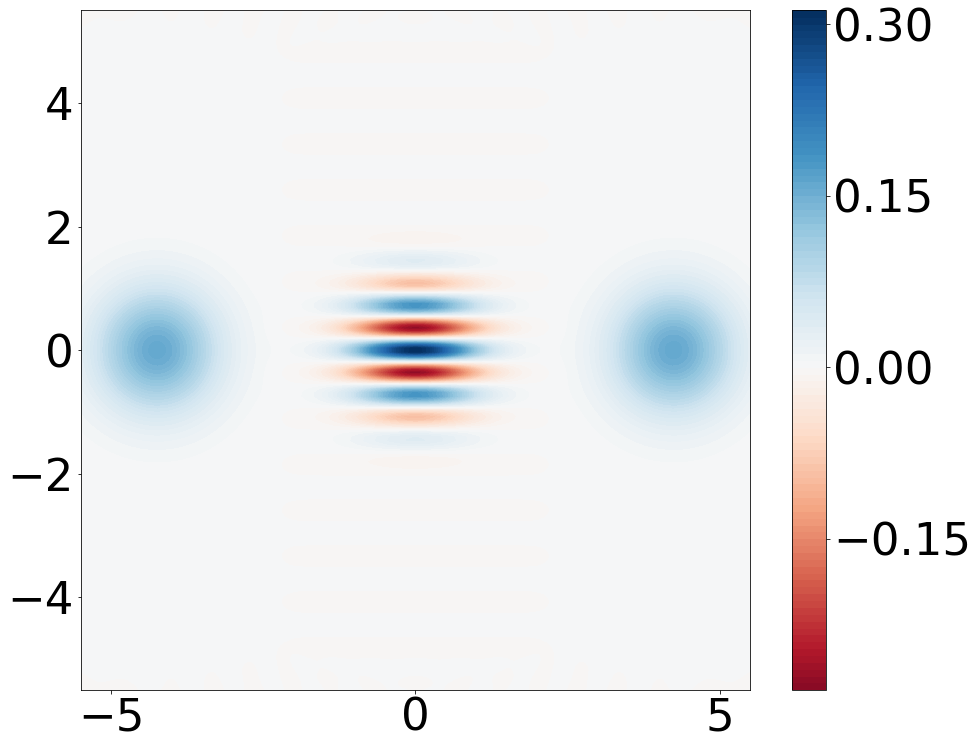}
    \caption{$\tau$ = 0}
\end{subfigure}
\hfill
\begin{subfigure}[t]{0.15\textwidth}
\centering
    \includegraphics[width = \textwidth]{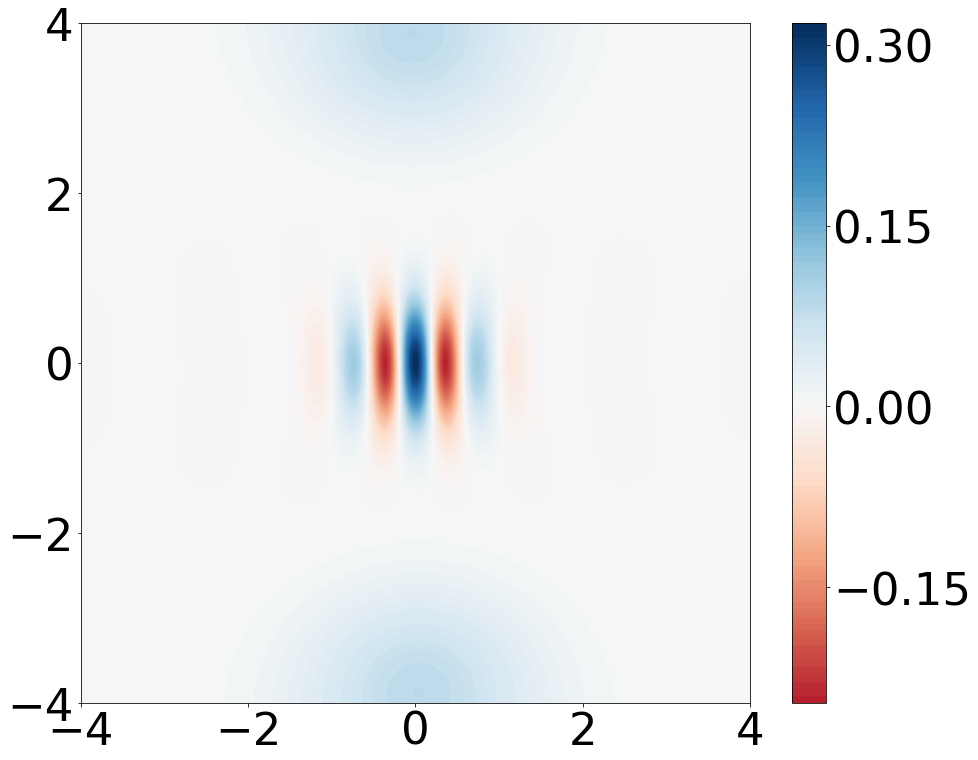}
    \caption{$\tau$ = $\frac{3\pi}{2}$ }
\end{subfigure}
\hfill
\begin{subfigure}[t]{0.155\textwidth}
\centering
    \includegraphics[width = \textwidth]{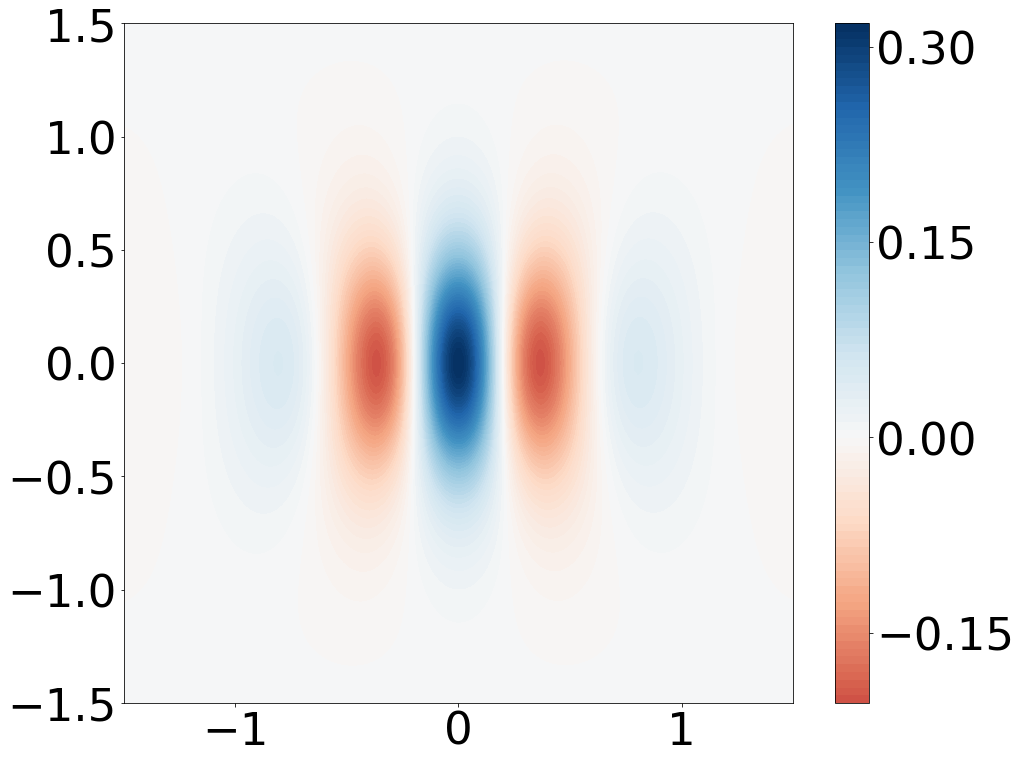}
    \caption{$\tau = \frac{9 \pi}{2}$}
\end{subfigure}
\begin{subfigure}[t]{0.15\textwidth}
\centering
    \includegraphics[width=\textwidth]{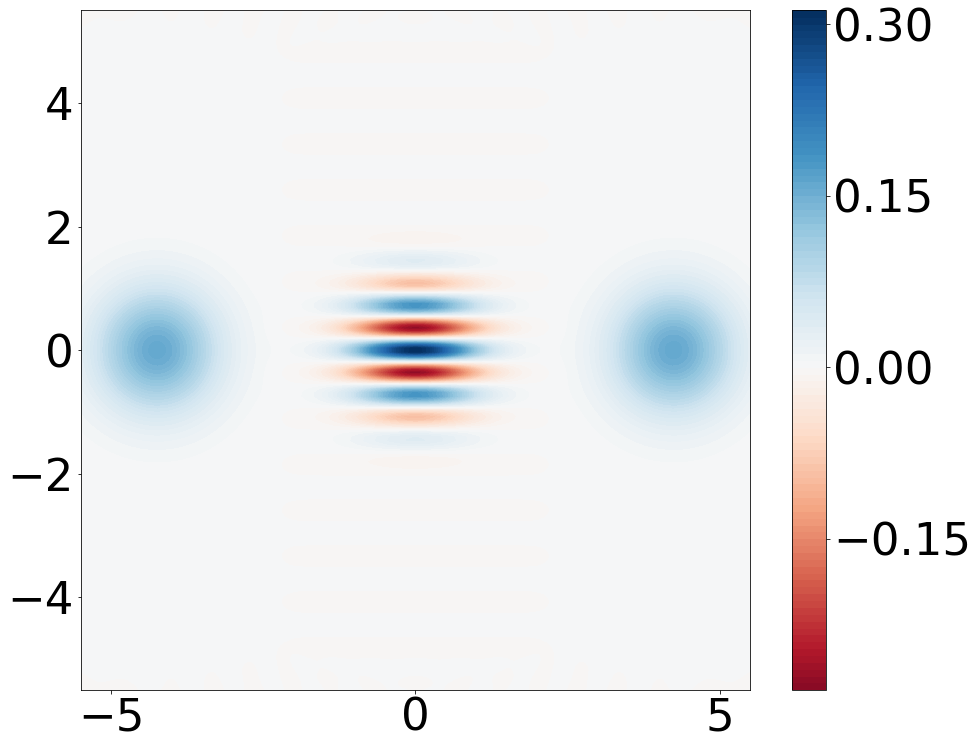}
    \caption{$\tau$ = 0}
\end{subfigure}
\hfill
\begin{subfigure}[t]{0.15\textwidth}
\centering
    \includegraphics[width=\textwidth]{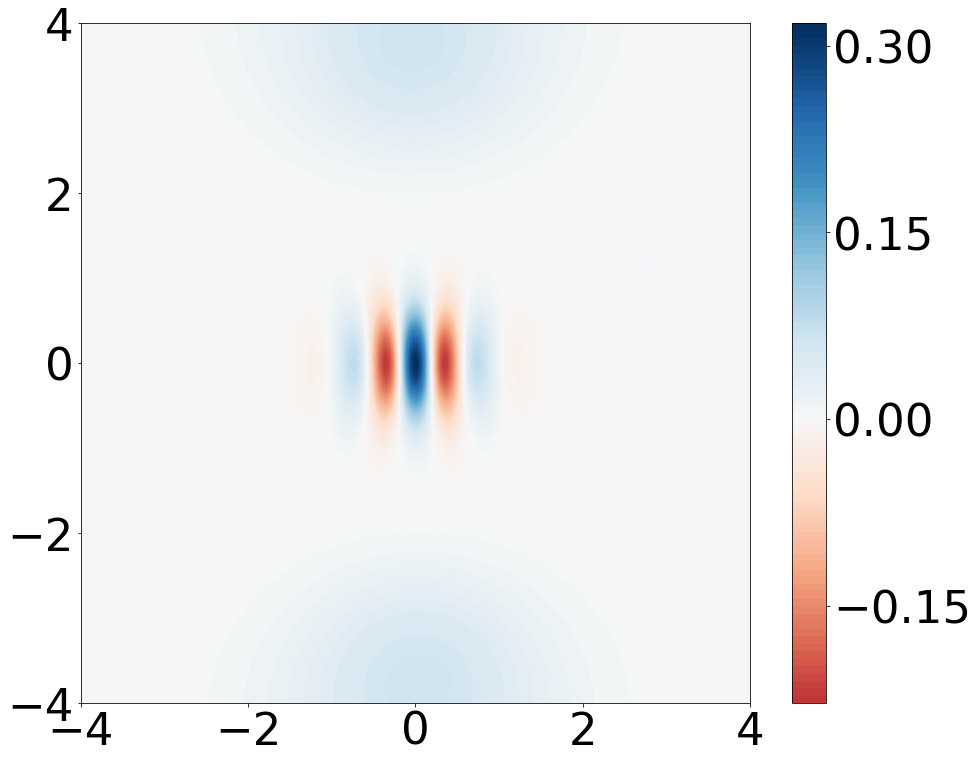}
    \caption{$\tau$ = $\frac{3\pi}{2}$}
\end{subfigure}
\hfill
\begin{subfigure}[t]{0.152\textwidth}
\centering
    \includegraphics[width=\textwidth]{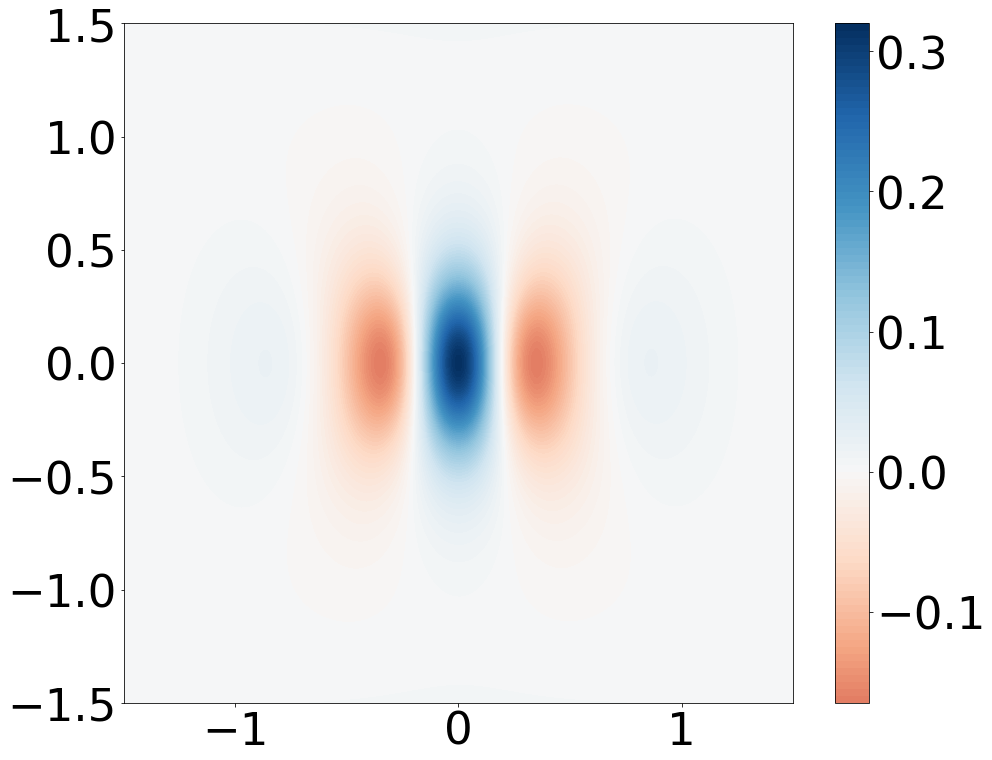}
    \caption{$\tau$ = $\frac{9\pi}{2}$}
\end{subfigure}
\caption{\label{wignerfig}Wigner function snapshots at different times. Horizontal coordinate is for dimensionless position $x\sqrt{M\Omega /\hbar}$ and vertical coordinate is for dimensionless momentum $p/\sqrt{M\Omega \hbar}$. The example parameters are (a), (b) and (c): $\alpha = 3$, $n = 3$ and $Q^{-1} = 0.001$; (d), (e) and (f): $\alpha = 3$, $n = 5$ and $Q^{-1} = 0.001$.}
\end{figure}

In the following, we consider the evolution of system oscillator initial coherent state superpositions of the form
\begin{align}
|\psi(0) \rangle = N\left(|\alpha\rangle + |-\alpha\rangle\right),
\label{supeq}
\end{align}
where $N$ is a normalization constant.
Figure \ref{wignerfig}  displays the evolving state through its Wigner function representation \cite{case2008} for a selection of $\alpha$, $n$, and $Q$ parameter values--obtained by numerically solving the master equation (\ref{quantummastereq}).
Quantum coherence manifested in the presence of negative-valued Wigner function regions can survive longer than the amplitude damping time. This is to be contrasted with the commonly-investigated quantum Brownian oscillator model with single photon damping, described by the following master equation:
\begin{align}
\frac{d \rho}{dt} =& {i\Omega}[\rho,a^\dag a] + \frac{\gamma}{2}(n+1)\left(2a\rho  a^\dag - a^\dag  a\rho\right. \nn \\
-&\left.\rho a^\dag a \right) 
+\frac{\gamma}{2}n \left(2a^\dag \rho a - aa^\dag \rho - \rho a  a^\dag \right).
\label{qbmeq}
\end{align}
For the latter master equation, decoherence proceeds more rapidly than amplitude damping. Note that the initial, even superposition state (\ref{supeq}) is an eigenstate of the operator $a^2$ since $a^2|\pm\alpha\rangle=\alpha^2|\pm\alpha\rangle$, so that the two-photon loss term in the master equation (\ref{quantummastereq}) preserves coherence \cite{gilles1993}. In contrast, the even superposition state (\ref{supeq}) flips to the odd superposition state $N\left(|\alpha\rangle - |-\alpha\rangle\right)$ under the action of a single annihilation operator $a$, hence the single photon loss term in the master equation (\ref{qbmeq}) does not preserve coherence.   

\begin{figure}[htbp]
    \centering
\begin{subfigure}[b]{0.21\textwidth}
\centering
\caption{}
    \includegraphics[width=\textwidth]{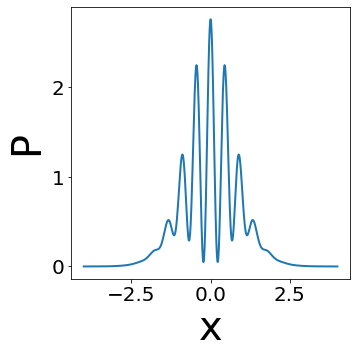}
\end{subfigure}
\hfill
\begin{subfigure}[b]{0.22\textwidth}
\centering
 \caption{}
    \includegraphics[width=\textwidth]{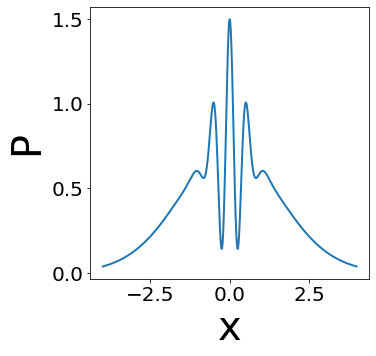}
\end{subfigure}

\begin{subfigure}[b]{0.22\textwidth}
\centering
 \caption{}
    \includegraphics[width=\textwidth]{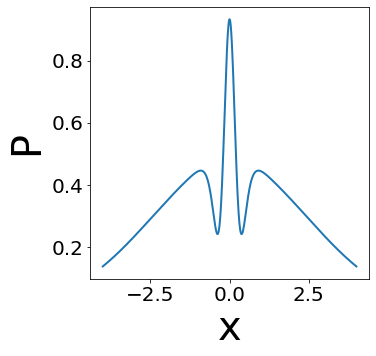}
\end{subfigure}
\hfill
\begin{subfigure}[b]{0.22\textwidth}
\centering
 \caption{}
    \includegraphics[width=\textwidth]{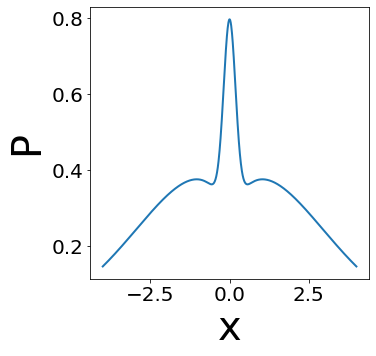}
\end{subfigure}
   \caption{\label{interferencefig}Snapshots of the (unnormalized)  position probability density $P$ vs dimensionless position coordinate $x\sqrt{M\Omega /\hbar}$ when the two initial coherent states in the  superposition pass through each other at $x=0$ at times (a) 
    $\tau=\pi/2$, (b) $\tau=(6+1/2)\pi/2$, (c) $\tau=(42+1/2)\pi/2$ and (d) $\tau=(190+1/2)\pi/2$ . The example parameters  are $Q^{-1} = 0.0005$, $\alpha = 5$,  and $n = 3$. The probability density should be understood with an overall normalization constant. }
\end{figure}
Figure \ref{interferencefig} gives snapshots of the system oscillator position probability density $P(x,t)=\langle x|\rho(t)|x\rangle$ when the two initial coherent state wavefunctions making up the superposition pass through each other at $x=0$ (at time instants $\tau_k=\Omega t_k={\pi}(k+1/2),\, k=0,1,2,\dots$). These snapshots can be interpreted as the marginal probability distributions obtained by integrating over the momentum coordinate of Wigner function distributions that are similar to those shown in Fig. \ref{wignerfig} (but for different parameter values). The presence of quantum coherence is manifested in $P(x,t)$ having an oscillatory dependence about $x=0$. In contrast to the gravity toy model (c.f., Fig. \ref{GRp_dis}), the interference fringes survive longer than the initial coherent state peaks; even after 190 cycles a small amount of interference is still present, while the initial coherent states have decayed away.

\begin{figure}[
htbp]
    \centering
\begin{subfigure}[h]{0.23\textwidth}
\centering
\caption{}
    \includegraphics[width=\textwidth]{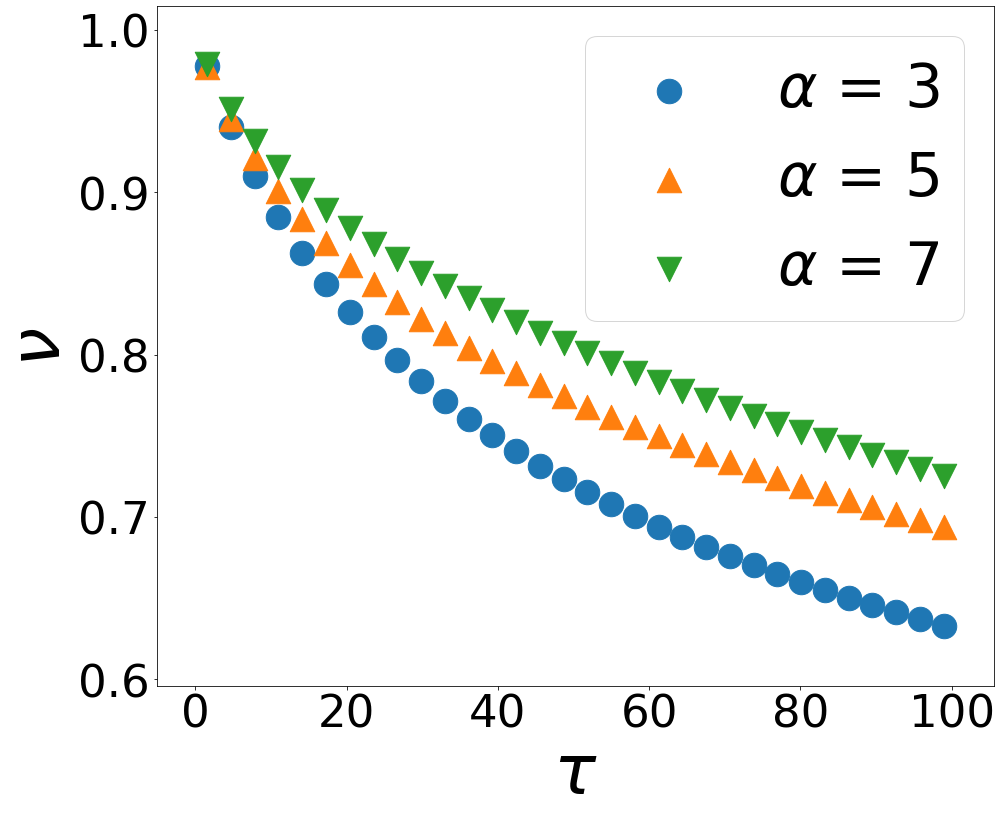}

\end{subfigure}
\vfill
\begin{subfigure}[h]{0.23\textwidth}
\centering
 \caption{}
    \includegraphics[width=\textwidth]{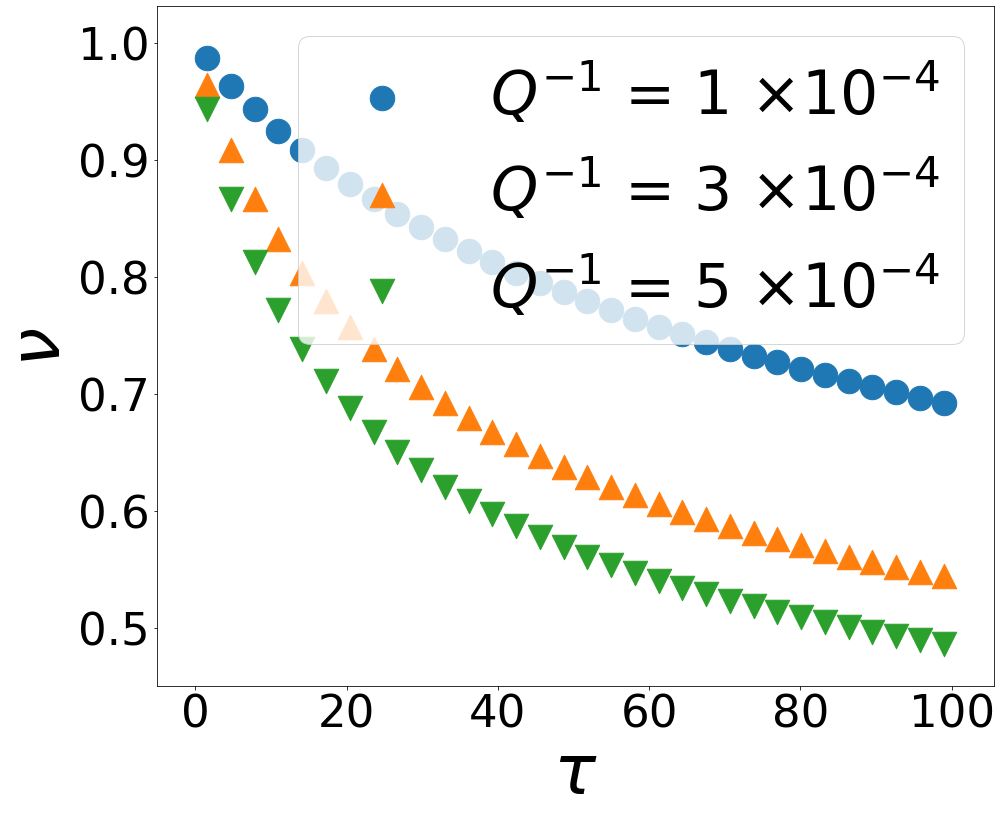}
   
\end{subfigure}
\begin{subfigure}[h]{0.23\textwidth}
\centering
 \caption{}
    \includegraphics[width=\textwidth]{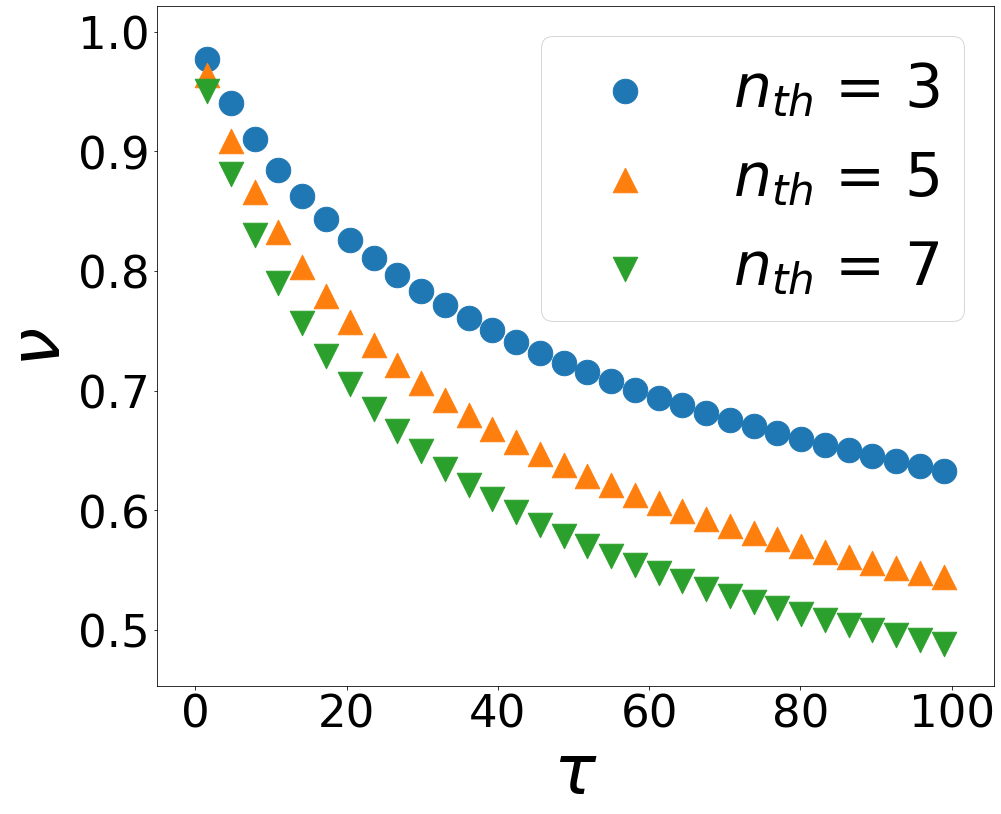}
   
\end{subfigure}
   \caption{\label{visibilityfig}Visibility as a function of dimensionless time $\tau=\Omega t$. The example parameters are (a) $Q^{-1}$ = 0.0003, $n = 3$; (b)   $\alpha$ = 3, $n = 5$; (c) $Q^{-1}$ = 0.0003, $\alpha = 3$.}
\end{figure}
Proceeding as in Sec. \ref{gravdecohsec} for the scalar gravity model, We can operationally quantify the decoherence of an initial superposition of coherent states by using the fringe visibility measure $\nu$ (\ref{visibilityeq}) for the position detection probability density.   Figure \ref{visibilityfig}  shows the time dependence of the visibility  $\nu$ for a range of parameter choices. The rate at which the visibility is reduced increases with larger damping parameter and bath temperature as for the single photon damping case with master equation (\ref{qbmeq}), but contrary to single photon damping   the visibility reduction rate decreases with larger initial amplitude.  

%========================================
%========================================

%========================================
%========================================
\section{Concluding Remarks}
\label{conclusionsec}
In the present work, we have explored two 0d system-bath models that share common features with a scalar field system weakly coupled to gravity, and also with scalar QED. The considered model systems comprise a single harmonic oscillator, with the gravitational and electromagnetic fields replaced by a bath of harmonic oscillators, in each case coupled to the oscillator system via  non-quadratic interaction terms that resemble the respective scalar-weak field gravity and scalar QED interactions. We utilized these models as a test bed for an operational interference fringe visibility measure of decoherence, as well as for various standard open quantum systems approximation methods.

In particular, we have gained several insights working with the two models that may be of use for analyzing gravitational decoherence: (1) A relatively straightforward, interferometric operational approach for verifying decoherence dynamics can be analyzed that does not involve just extracting the off-diagonal terms of the system reduced density matrix (which is not a gauge invariant quantity in the full theory). (2) While the full scalar matter-gravitational system likely cannot be solved exactly as is the case for the corresponding model system, verified standard open quantum systems approximation methods (e.g., deriving a RWA quantum Langevin equation) may be applicable to the full system; the more involved closed time path integral approaches that are commonly applied to such dynamical quantum field system problems \cite{calzetta2008} can be guided by the simpler approximation method approaches that are common to non-relativistic open quantum systems analyses.   (3) `Planckian', cut-off dependent terms can affect the initial decoherence dynamics. However, by being careful with the choice of initial system-environment state taking into account finite state preparation times, such cut-off dependence may be avoided.  

The logical next step will be to apply  the considered approximation methods to the scalar QED system, and verify that the interferometric observable quantities for probing decoherence are gauge invariant and accessible to analysis at  low energy (i.e., `table top' experiment) scales. We can then apply the lessons learned from the 0d models as well as the full scalar QED model to the scalar matter--weak gravity system. While the latter quantum field system is of course more challenging to analyze, the insights gained from the present work might nevertheless serve as a useful guide in developing an operational understanding of gravitational decoherence, just as the 0d model (\ref{0dgraveq}) proved valuable for the initial field theoretic investigation in Ref. \cite{blencowe2013}.

\begin{acknowledgments}
 We thank Sougato Bose and William Braasch for very helpful discussions. This work was supported in part by the NSF under Grant Nos. PHY-2011382 and  DMR-150738.
\end{acknowledgments}

\bibliography{main}

\end{document}